\RequirePackage{ifpdf}
\documentclass[article]{JHEP3}
\pdfoutput=1
\usepackage{amsmath}

\usepackage{amssymb}
\usepackage{graphicx}
\usepackage{subfig}

\usepackage{amsfonts}

\def\beq{\begin{equation}}
\def\eeq{\end{equation}}
\def\bea{\begin{eqnarray}}
\def\eea{\end{eqnarray}}
\def\nn{\nonumber}
\def\f{\frac}

\title{  \begin{center}   
The Superconformal Index of  \\ Class ${\cal S}$ Theories of~Type~$D$
\end{center}
}
\preprint{YITP-SB-12-45}
\author{Madalena Lemos\footnote{madalena.lemos@stonybrook.edu}\;,
$\,$Wolfger Peelaers\footnote{wolfger.peelaers@stonybrook.edu}\;,
$\,$Leonardo Rastelli\footnote{leonardo.rastelli@stonybrook.edu} 
\\
\\
\it C.N. Yang Institute for Theoretical Physics,\\
\it Stony Brook University, \\
\it Stony Brook, NY 11794-3840, USA
}
\abstract{We consider
the superconformal index of class ${\cal S}$ theories of type $D$, which arise by compactification
of the $(2,0)$ $D_n$ theories on a punctured Riemann surface ${\cal C}$. We also allow for
the presence of twist lines on ${\cal C}$ associated to the $\mathbb{Z}_2$ outer automorphism of $D_n$.
For the two-parameter slice $(p=0, q, t)$ in the space of superconformal fugacities, we determine the 
$2d$ TQFT that computes the index.  
 }

\begin{document}
%\maketitle \flushbottom

%\tableofcontents
%=====================================%
\section{Introduction}
%=====================================%

Class ${\cal S}$ theories are a family of four-dimensional gauge theories with ${\cal N}=2$ supersymmetry,
which arise by (partially twisted) compactification of the six-dimensional $(2,0)$ theories on
 a punctured Riemann surface \cite{Gaiotto:2009we, 2009arXiv0907.3987G}.
 There is a beautiful dictionary relating supersymmetric observables of 
  the $4d$ theory with quantities defined on the surface ${\cal C}$. Notably,
  the complex structure moduli of ${\cal C}$ correspond to the exactly marginal $4d$ gauge couplings,
  while the punctures are associated to flavor symmetries. The partition function of the $4d$ theory on $S^4$ is computed by a conformal field theory
correlator on ${\cal C}$ \cite{agt}.
Here we  focus on the item of the $4d/2d$ dictionary introduced in \cite{Gadde:2009kb}:
the $S^3 \times S^1$ partition function of a superconformal theory
of class ${\cal S}$, also known as the superconformal index \cite{Kinney:2005ej, 2006NuPhB.747..329R} (henceforth simply the index), is computed by a topological
QFT (TQFT) correlator on ${\cal C}$.

The $(2,0)$ theories are isolated superconformal field theories labeled by the simply laced Lie algebras, $\{ A_n, D_n, E_6, E_7, E_8 \}$.
Correspondingly, there are class ${\cal S}$ theories of type $A$, $D$ and $E$. 
To characterize the $4d$ theory one needs to further specify the punctured surface ${\cal C}$, together with some extra
discrete data associated to each puncture, which determine the flavor group associated to the puncture \cite{Gaiotto:2009we, Tachikawa:2009rb,Tachikawa:2010vg, DistlerA, 2011arXiv1106.5410C,DistlerTachikawa}.
This construction can be further enriched \cite{Tachikawa:2009rb,Tachikawa:2010vg} by 
decorating ${\cal C}$ with topologically non-trivial  ``twist lines''  (ending at a puncture or wrapping a cycle),
associated to the outer automorphism group of the simply laced Lie algebra (which is $\mathbb{Z}_2$ in all cases except $D_4$,
when it is $\mathbb{Z}_3$).

For all theories of type $A$, and in the absence of twist lines,
the superconformal index has been completely determined in  a series of papers \cite{Gadde:2009kb,2010JHEP...08..107G,2011PhRvL.106x1602G,Gadde:2011uv,Gaiotto:2012xa}\footnote{See also \cite{2012JHEP...05..007G, 2012JHEP...10..187G} for the evaluation of the index in the presence of BPS line defects and domain walls.},
by characterizing  the associated $2d$ TQFT. The TQFT is defined abstractly in terms
of its structure constants  $C_{\alpha \beta \gamma}$ (corresponding to three-punctured spheres) 
 and propagators $\eta^{\alpha \beta}$ (corresponding to  two-punctured spheres), where $\alpha, \beta, \gamma$ label $A_n$
 irreducible representations.
For a two-dimensional slice $(p=0, q, t)$ in fugacity space, the answer takes an elegant closed form
involving  Macdonald polynomials, and the TQFT is recognized as $q$-deformed $2d$ Yang-Mills \cite{2005NuPhB.715..304A} in the zero area limit\footnote{See \cite{2012PhLB..716..450K, 2012arXiv1210.2855F} for a recent top-down argument that recovers $2d$ $q$YM by localization of $5d$ super Yang-Mills on $S^3$.} for $q=t$,
and as a certain refinement thereof~for~{$q \neq  \nobreak t$} \cite{2011arXiv1105.5117A,  2012arXiv1210.2733A}. This  result was originally found in \cite{2011PhRvL.106x1602G, Gadde:2011uv} by
focussing on the $A_1$ theories, which have a Lagrangian description, and finding
a basis of functions 
where the structure constants are diagonal
({\it i.e.}, $C_{\alpha \beta \gamma} = 0$ unless $\alpha = \beta = \gamma$). Since these functions are closely related
to Macdonald polynomials, which are defined for any root system,  a general answer can be naturally conjectured
for all $A_n$ theories \cite{Gadde:2011uv}.  

The conjecture of \cite{Gadde:2011uv} was recently proved in \cite{Gaiotto:2012xa}, and in fact extended to arbitrary $(p, q, t)$,
under the sole assumption
that class ${\cal S}$ theories enjoy generalized
 S-duality, which is the statement that  different pairs-of-pants decomposition
 of ${\cal C}$ correspond to the same $4d$ theory in different duality frames. Schematically, the strategy of \cite{Gaiotto:2012xa} was
 to derive certain difference equations for the index, by considering
 its singularity structure (the residues at the flavor fugacities poles) in distinct, but by assumption equivalent, duality frames.
 These difference equations have unique solutions and thus completely characterize the index. In particular, the
 eigenfunctions $\{ \psi_\alpha ( {\bf a} ) \}$ of the difference operators define the basis where the structure constants $C_{\alpha \beta \gamma}$
are diagonal.
 For arbitrary
 $(p, q, t)$, the difference operators are closely related to the elliptic RS operators, whose eigenfunctions are not known in closed form,
 but for $p=0$ they are related to the well-known Macdonald operators,
 whose eigenfunctions are the Macdonald polynomials.
Acting with a difference operator on the index has the physical interpretation \cite{Gaiotto:2012xa} of decorating the
$4d$ theory by the insertion of a BPS surface defect.

In this paper we enlarge the setup to include  class ${\cal S}$ theories of type $D$, also allowing for the possibility of $\mathbb{Z}_2$ twist lines on ${\cal C}$.
In type $D$ theories, twist lines are very natural, 
indeed they are necessary for the description of all the non-trivial examples with a Lagrangian, such
as the superconformal linear quivers with alternating $\mathrm{SO}/\mathrm{USp}$ gauge groups \cite{Tachikawa:2009rb,Tachikawa:2010vg}.
Ideally, one would generalize the approach of \cite{Gaiotto:2012xa}, and derive the general answer with no guesswork.
However this seems technically challenging, because conformal tails of type $D$ have no analog
of the $U(1)$ flavor punctures  that were used in \cite{Gaiotto:2012xa} to derive the difference operators.
Thinking about the physics of surface operators may provide the right clues, but we leave this for the future.
Here we generalize instead the approach of \cite{Gadde:2011uv}, and look  for a diagonal representation of the index for $p=0$
in the Macdonald basis. The presence of twist lines makes the story richer,
leading to an interesting extension of  $2d$ TQFT structure. As in \cite{Gaiotto:2012xa}, we get some mileage
by considering the action of Macdonald difference operators. In particular
we use these difference operators to argue that the index of free hypermultiplets has a diagonal expansion in the Macdonald basis.
In this paper the difference operators serve an auxiliary  technical role, but it is natural to expect
that they also have  a physical interpretation in terms of
BPS surface defects.

The rest of the paper is organized as follow. In section 2 we review some facts about the class ${\cal S}$ theories
of type $D$, with or without $\mathbb{Z}_2$ twist lines. In section 3, after briefly recalling the definition of the superconformal index
and the TQFT approach, we consider the special cases of $D_2$ and $D_3$ theories, and  extrapolate from them our general
proposal for the $D_n$ case (with maximal and empty punctures). In section 4 we consider the extension to partially closed punctures.
In appendix A we collect some technical background material on Macdonald theory. Finally  appendix B contains the main intertwining identity
that  shows diagonality of the free hyper index  in the Macdonald basis.

\medskip

{\it \noindent Note added:} Last night,  the interesting article \cite{2012arXiv1212.0545M} appeared on the ArXiv.
The authors of \cite{2012arXiv1212.0545M} focus on the evaluation
of the index twisted\footnote{For the case of ${\cal N}=4$ SYM, such twisted index was also studied in \cite{2012JHEP...01..116Z}.
The compactification of the $(2,0)$ theory on a circle with  automorphism twist was studied in \cite{2011JHEP...11..123T}.
} by  the outer automorphism group along the temporal $S^1$, while we
focus instead on the ordinary (untwisted) index of type $D$ theories but with twist lines on ${\cal C}$, so our results are largely complementary to theirs.
There is some partial overlap in the discussion of the purely $\mathrm{SO}$ theories.

%
%=====================================%
\section{Class ${\cal S}$ Theories of Type $D$}\label{Sec:Dtheories}
%=====================================%
%
The superconformal theories that we consider were constructed in~\cite{Tachikawa:2009rb, Tachikawa:2010vg}
following~\cite{Gaiotto:2009we}. 
Let us briefly
summarize the relevant points.

First recall that an $\mathcal{N}=2$ hypermultiplet in
representation $R$ of the gauge group $G$ can be decomposed in two
$\mathcal{N}=1$ chiral multiplets sitting in complex conjugate
representations $R$ and $R^*.$ The flavor symmetry of $N_f$ such hypermultiplets depends on the reality properties of the
representation $R.$ For $R$ complex one has flavor symmetry
$\mathrm{U}(N_f);$ for $R$ real, the flavor symmetry is enhanced
to $\mathrm{USp}(2N_f)$; for $R$ pseudoreal the flavor symmetry
is enlarged to $\mathrm{SO}(2N_f).$ For pseudoreal $R,$
there is no need to double the multiplet to satisfy CPT, and one can
consider a single chiral multiplet in representation $R$ as an
$\mathcal{N}=2$ multiplet. This is the so-called
half-hypermultiplet. However, to avoid Witten's global anomaly \cite{Witten} one cannot have an odd number of half-hypermultiplets. $N_f$ half-hypermultiplets have
$\mathrm{SO}(N_f)$ flavor symmetry.

In order to construct superconformal theories, all couplings need to
be marginal. For gauge group $\mathrm{SO}(m)$ with hypermultiplets
in the vector representation this implies that $N_f = m-2,$ and
then the flavor symmetry is $\mathrm{USp}(2m-4).$ For gauge group
$\mathrm{USp}(2n)$ with hypermultiplets in the fundamental
representation the requirement is that $N_f = 2n+2,$ and the flavor
symmetry is given by $\mathrm{SO}(4n+4).$ It is possible to consider
half-hypermultiplets in this case.

Given these constraints, one can start constructing superconformal
quiver gauge theories. In order to be superconformal, these will in
general have alternating gauge groups. Let us briefly introduce our
quiver conventions which follow \cite{Tachikawa:2009rb}. A gauge
group is depicted by a circle, grey when $\mathrm{SO}$ and black for
$\mathrm{USp}.$ Numbers inside the circles indicate which
$\mathrm{SO}$ or $\mathrm{USp}$ gauge group is considered. Lines
connecting two gauge groups represent bifundamental\footnote{More
precisely, the word ``fundamental'' means the vector representation of
$\mathrm{SO}$ and the fundamental representation of $\mathrm{USp}$}
half-hypermultiplets. Additional hypermultiplets are denoted by
squares. Grey squares indicate that they carry $\mathrm{SO}$ flavor
symmetry, and black ones $\mathrm{USp}$ flavor symmetry. Numbers
inside the boxes indicate which flavor symmetry they have.
\begin{figure}
\centering
\subfloat[]{\label{QuiverSOgauge2}\includegraphics[height=1cm]{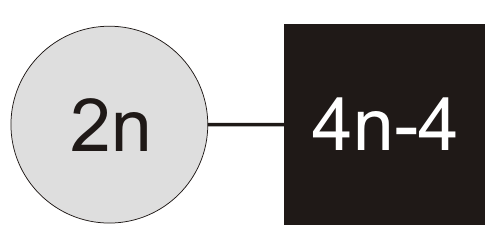}}
\quad \quad \quad
\subfloat[]{\label{QuiverSOgauge}\includegraphics[height=1cm]{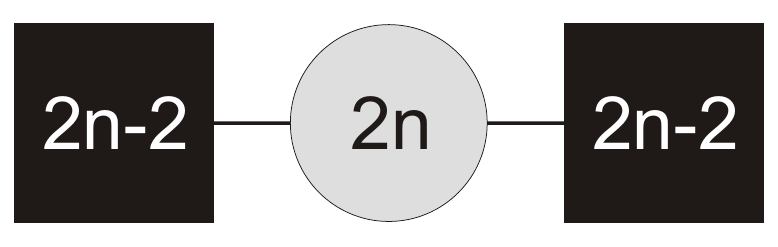}}
\quad \quad \quad
\subfloat[]{\label{GaiottoSOgauge}\includegraphics[height=1.3cm]{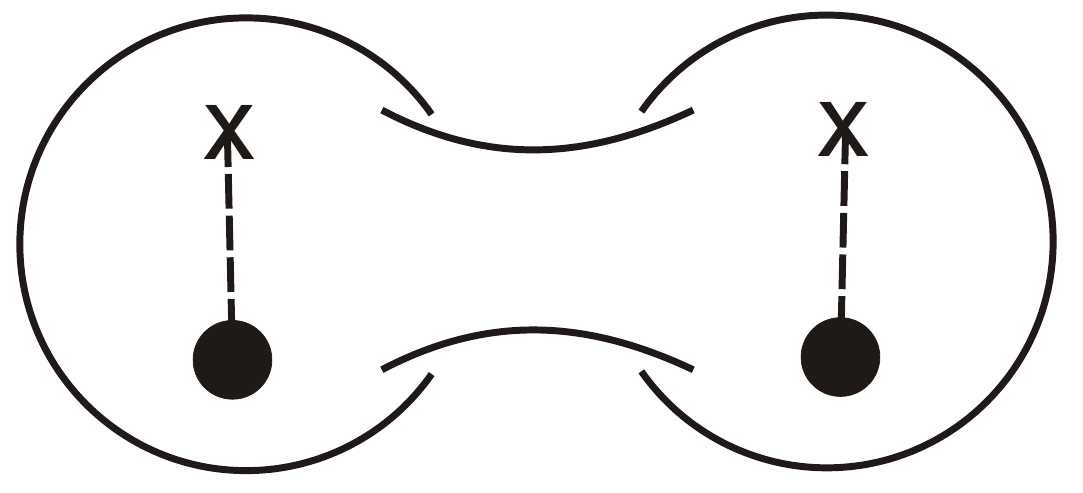}}
\quad \quad \quad
\caption{Quiver and corresponding curve for the
$\mathrm{SO}(2n)$ gauge theory with $N_f = 2n-2$ hypermultiplets in
the vector representation of $\mathrm{SO}(2n)$.}\label{SOgauge}
\end{figure}
The simplest example is to consider an $\mathrm{SO}(2n)$
gauge theory coupled to $N_f = 2n-2$ hypermultiplets. The flavor
symmetry is $\mathrm{USp}(4n-4).$ This theory is depicted in
figure~\ref{QuiverSOgauge2}. In figure~\ref{QuiverSOgauge} one focuses
on a $\mathrm{USp}(2n-2) \times \mathrm{USp}(2n-2)$ subgroup of this
flavor symmetry group. This quiver theory can also be depicted in
terms of the curve $\mathcal{C}$ in figure~\ref{GaiottoSOgauge}. Here the
$\mathrm{SO}$ gauge group is depicted by a cylinder and each three-punctured sphere
represents a free half-hypermultiplet transforming under the
fundamental of $\mathrm{USp(2n-2)},$ corresponding to the puncture
denoted by the symbol $\bullet$, and the vector of
$\mathrm{SO(2n)}$, depicted by the symbol $\circ$. In this case the
diagonal $\mathrm{SO}$ group is gauged. Also notice the presence of a $\mathbb{Z}_2$ twist line connecting the $\mathrm{USp}$ puncture and the empty
$\mathrm{USp}$ flavor symmetry puncture (drawn as
$\mathbf{\times}$), in the curve.
\begin{figure}
\centering
\subfloat[]{\label{QuiverUSpgauge2}\includegraphics[height=1cm]{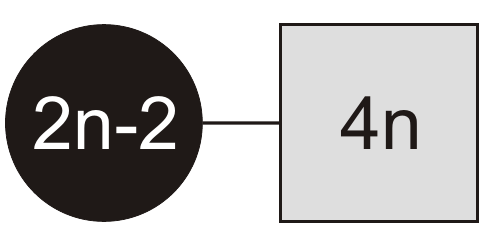}}
\quad \quad \quad
\subfloat[]{\label{QuiverUSpgauge}\includegraphics[height=1cm]{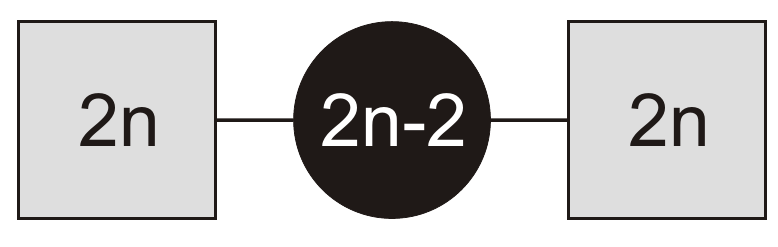}}
\quad \quad \quad
\subfloat[]{\label{GaiottoUSpgauge}\includegraphics[height=1.3cm]{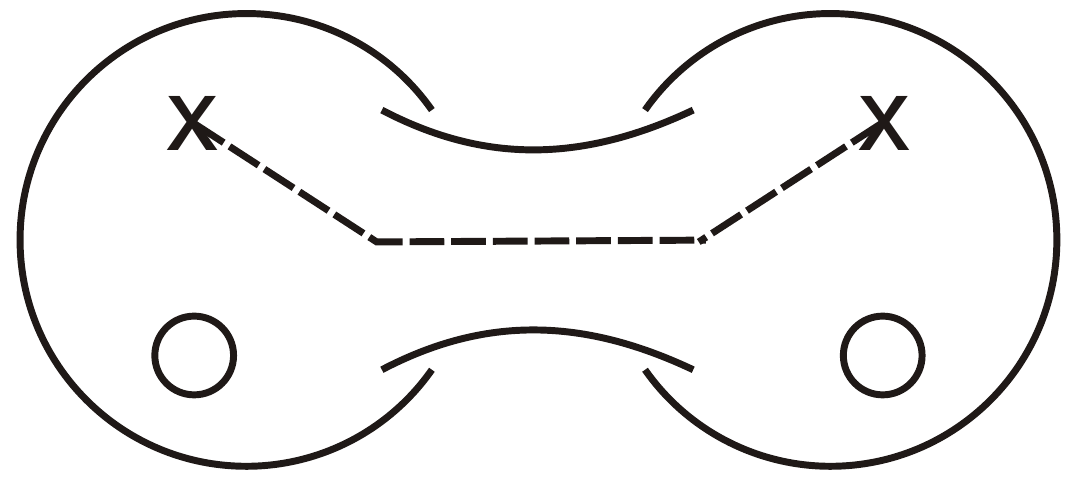}}
\quad \quad \quad
\caption{Quiver and corresponding curve for the
$\mathrm{USp}(2n-2)$ gauge theory with $N_f = 2n$ hypermultiplets in
the fundamental representation of
$\mathrm{USp}(2n-2)$.}\label{USpgauge}
\end{figure}
Similarly one can consider a $\mathrm{USp}(2n-2)$ gauge theory
coupled to $N_f = 2n$ hypermultiplets. Figure~\ref{USpgauge} depicts
this theory. Here the $\mathrm{USp}$ gauge group is denoted by a
cylinder with a twist line.

\begin{figure}
\centering
\label{TSOlinearquiver}\includegraphics[width=12cm]{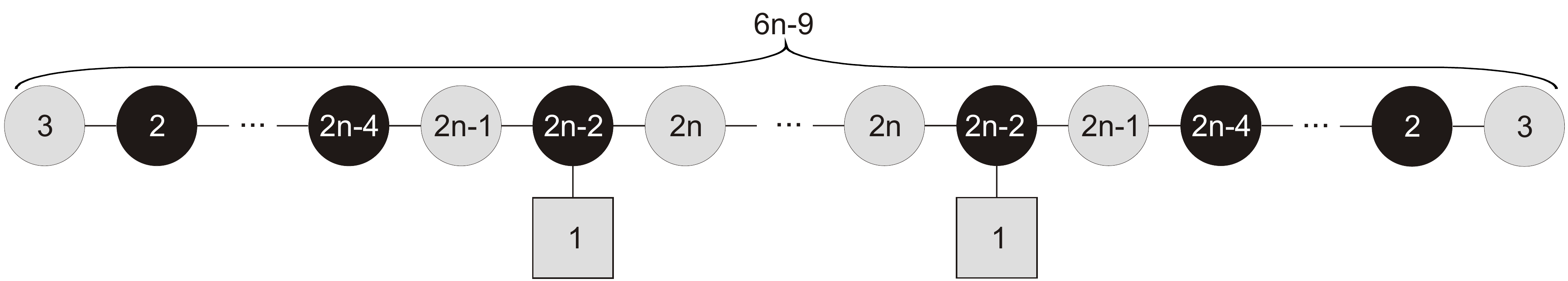}\\
\label{TSOfinal}\includegraphics[width=8cm]{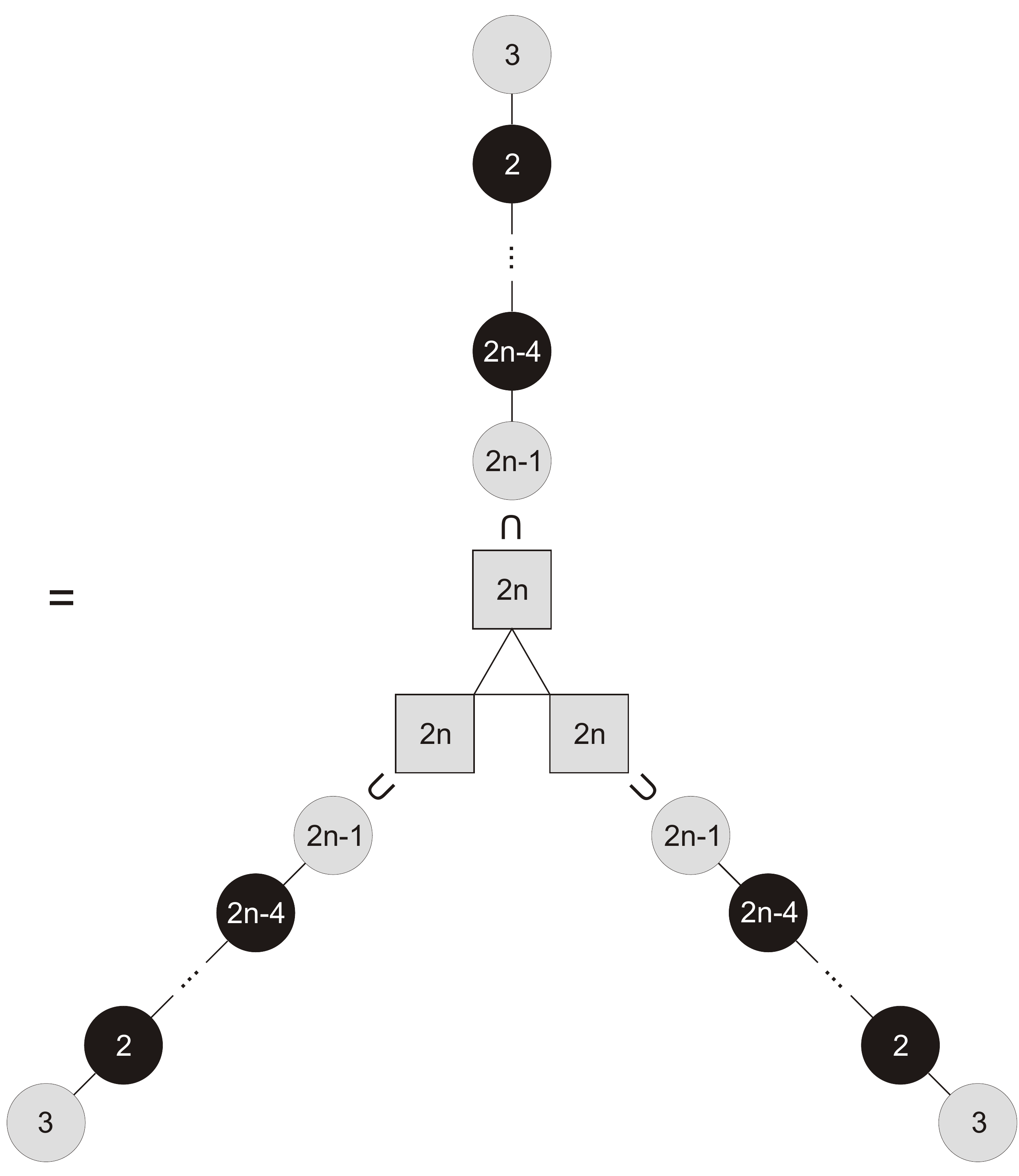}
\caption{Construction of the $T_{\mathrm{SO}(2n)}$ theory. The
$\cdots$ in the middle of the top quiver stand for a sequence of
alternating $\mathrm{SO}(2n)$ and $\mathrm{USp}(2n-2)$ gauge groups.
The other $\cdots$ on the left/right stand for a sequence of alternating odd
$\mathrm{SO}$ and $\mathrm{USp}$ gauge groups of increasing/decreasing rank. The symbol $\subset$ means that we gauge a $\mathrm{SO}(2n-1)$ subgroup of $\mathrm{SO}(2n)$.}
\label{TSO}
\end{figure}
Starting with linear quivers, which have a Lagrangian description, one can
make use of dualities to obtain new interacting theories.
Figure~\ref{TSO} summarizes the construction of the so-called
$T_{\mathrm{SO}(2n)}$ theory~\cite{Tachikawa:2009rb}. It is
described by a three-punctured sphere with three $\mathrm{SO}(2n)$
punctures. The effective number of hyper- and vectormultiplets for
this theory is~\cite{Tachikawa:2009rb}
\beq n_V=\f{8 n^3}{3}-7 n^2+\f{10 n}{3}\,,\quad
n_H=\f{8 n^3}{3}-4 n^2+\f{4 n}{3}\,.\label{nVTSO}
\eeq
Notice that for $n=2$ one finds that the effective number of
vectormultiplets is zero.
\begin{figure}
\centering
\label{TSOlinearquivertilde}\includegraphics[width=12cm]{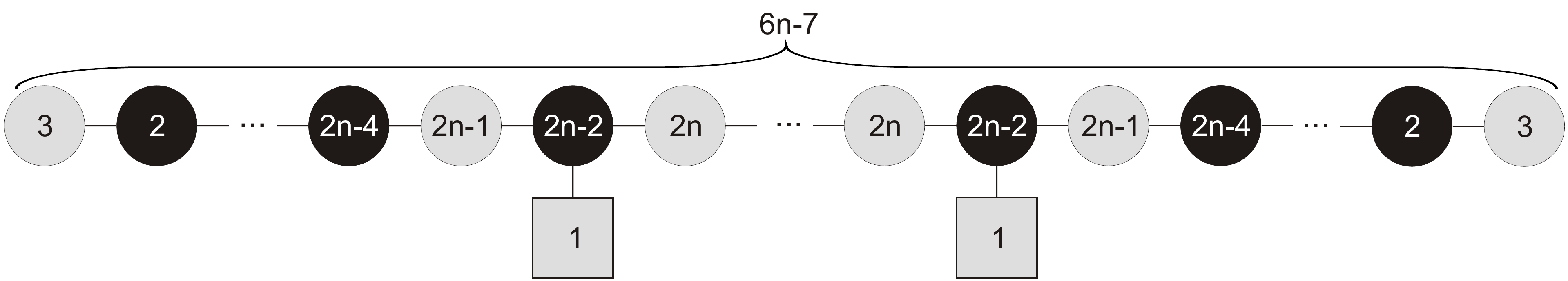}\\
\label{TSOtildefinal}\includegraphics[width=8cm]{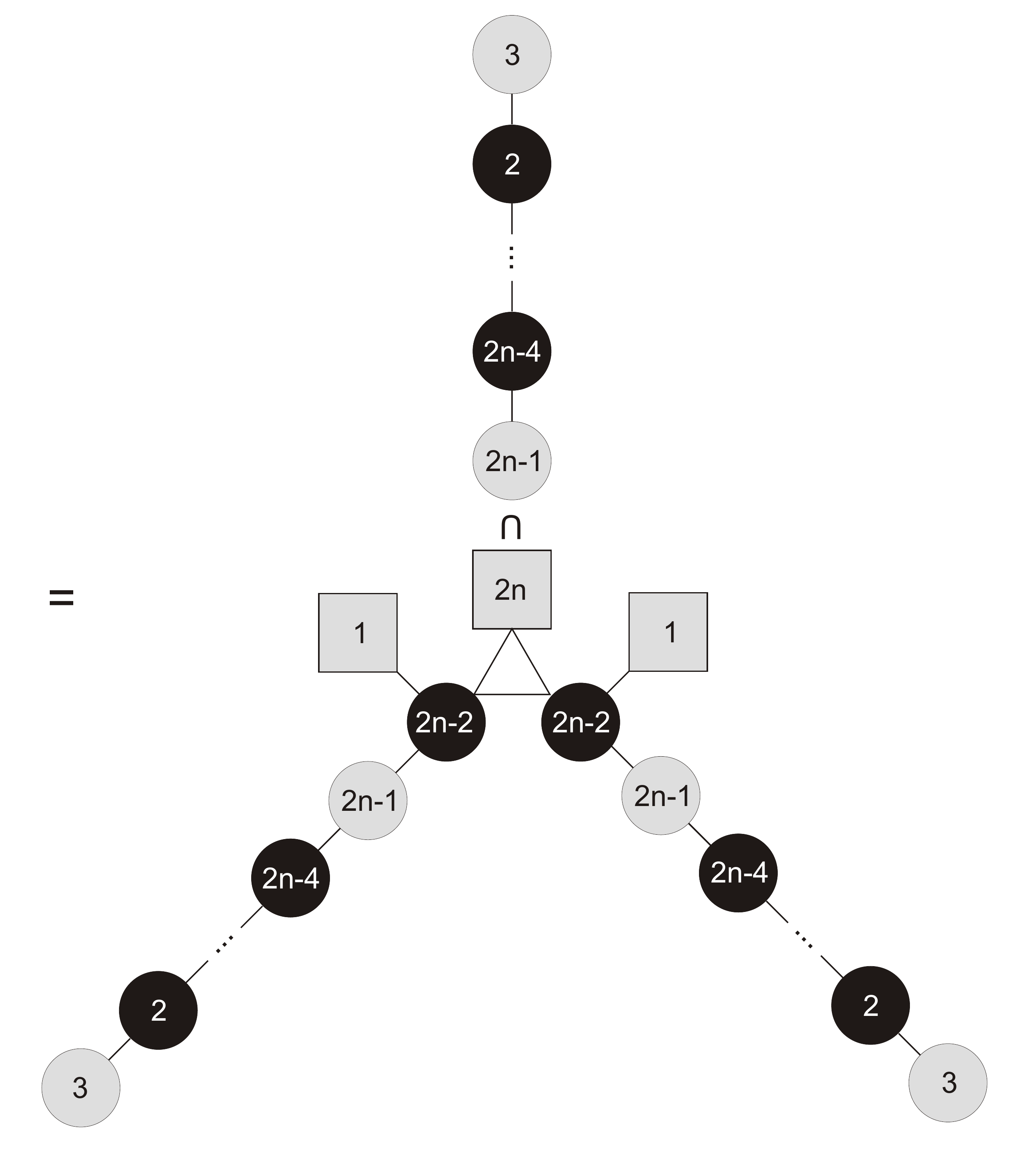}

\caption{Construction of the $\tilde T_{SO(2n)}$ theory.
The $\cdots$ in the
middle of the top quiver stand for a sequence of alternating
$\mathrm{SO}(2n)$ and $\mathrm{USp}(2n-2)$ gauge groups. The other
$\cdots$ on the left/right stand for a sequence of alternating odd $\mathrm{SO}$ and
$\mathrm{USp}$ gauge groups of increasing/decreasing rank. The symbol $\subset$ means that we gauge a $\mathrm{SO}(2n-1)$ subgroup of $\mathrm{SO}(2n)$.}\label{TSOtilde}
\end{figure}
Figure~\ref{TSOtilde} summarizes the construction of another
interacting theory~\cite{Nishinaka:2012vi}, denoted as
$\tilde{T}_{\mathrm{SO}(2n),}$ which is described by a sphere with
two $\mathrm{USp}(2n-2)$ punctures and one $\mathrm{SO}(2n)$
puncture. The effective number of hyper- and vectormultiplets for
this theory is~\cite{Nishinaka:2012vi}
\beq \label{nvhtilde} n_V=\f{8 n^3}{3}-7 n^2+\f{16 n}{3}-1\,,\quad
n_H=\f{8 n^3}{3}-4 n^2+\f{4 n}{3}\,.
\eeq

Finally, there is one more basic three-punctured sphere we can introduce and that
will only appear as part of a bigger theory, namely a sphere with an
$\mathrm{SO}(2n)$ puncture and two empty $\mathrm{USp}$ punctures
$\mathbf{\times}$.
\begin{figure}
\centering
$\mathcal{I}(\mathbf{a},\mathbf{b})= \vcenter{\hbox{\subfloat[]{\label{sphereC}\includegraphics[width=1.5cm]{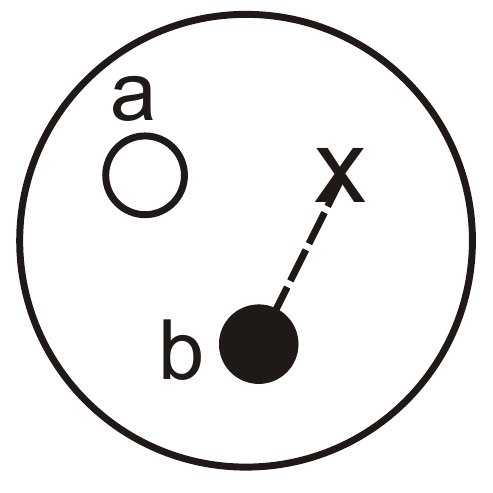}}}}$\,, \quad \quad
$\mathcal{I}_{T_{\mathrm{SO}(2n)}}(\mathbf{a}_1,\mathbf{a}_2,\mathbf{a}_3)= \vcenter{\hbox{\subfloat[]{\label{sphereB}\includegraphics[width=1.5cm]{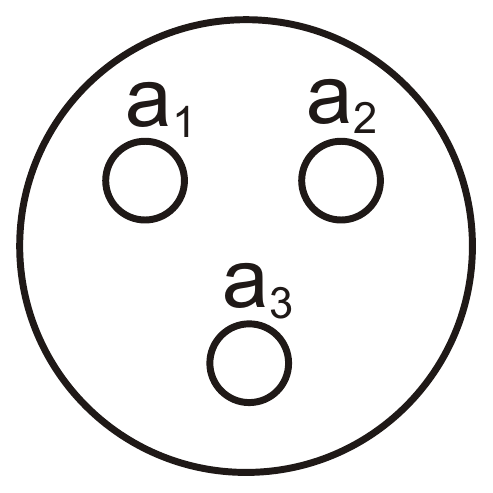}}}}$\,, \\
$\mathcal{I}(\mathbf{a})= \vcenter{\hbox{\subfloat[]{\label{sphereE}\includegraphics[width=1.5cm]{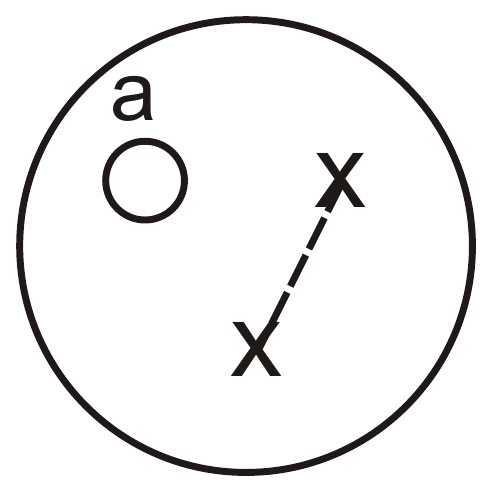}}}}$\,,\quad \quad
$\mathcal{I}_{\tilde{T}_{\mathrm{SO}(2n)}}(\mathbf{a},\mathbf{b}_1,\mathbf{b}_2)= \vcenter{\hbox{\subfloat[]{\label{sphereA}\includegraphics[width=1.5cm]{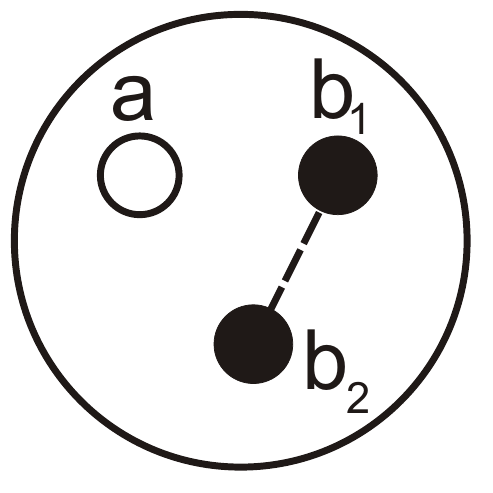}}}}$ .
\caption{The three-punctured spheres with maximal $\mathrm{SO}$ punctures and maximal or empty $\mathrm{USp}$ punctures.
 We indicate the flavor fugacity
assignments and our notation for the corresponding
index.}\label{basicspheres}
\end{figure}

\begin{figure}
\centering
\subfloat[]{\label{Sdual1}\includegraphics[width=7cm]{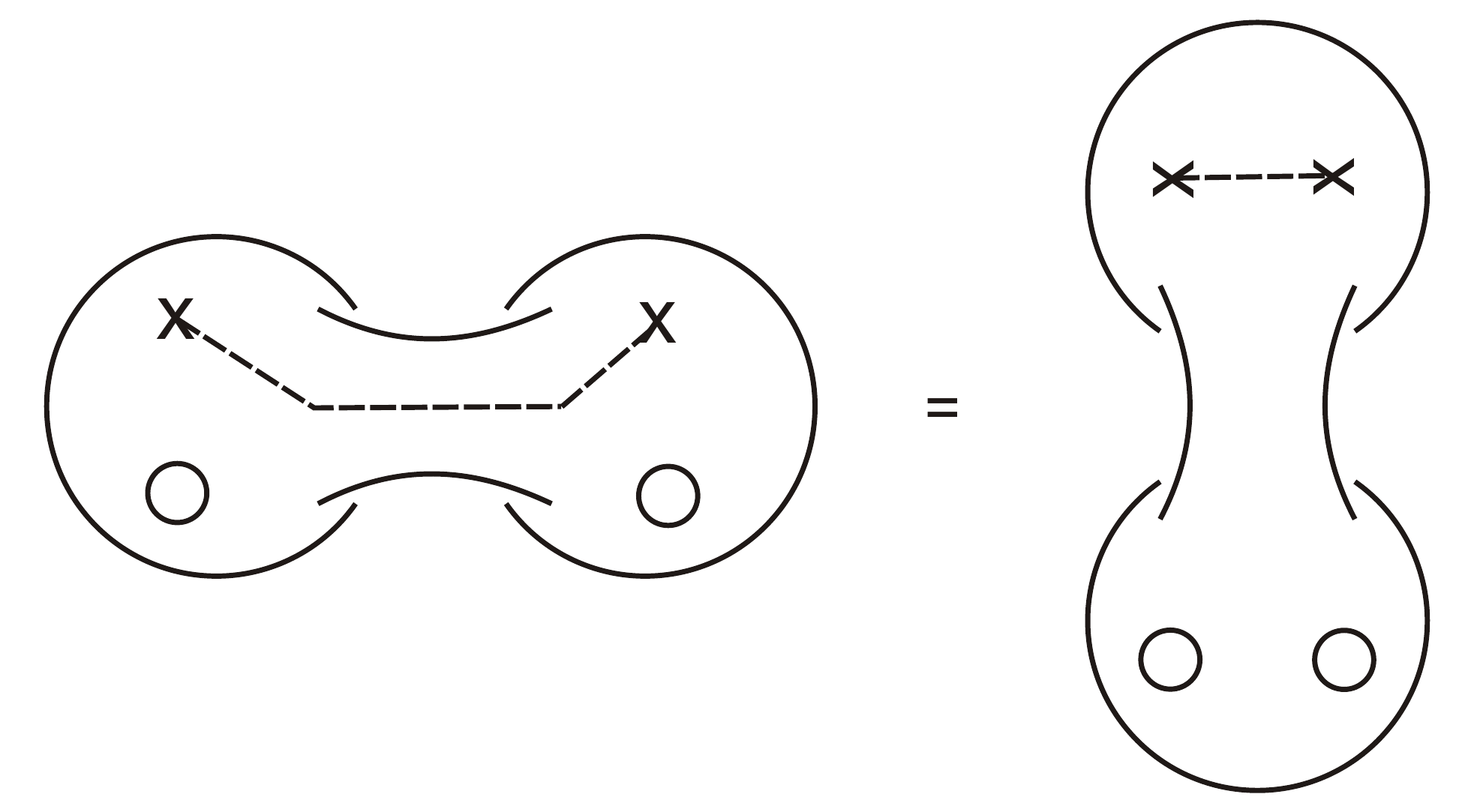}}    \quad \quad \quad
\subfloat[]{\label{Sdual2}\includegraphics[width=7cm]{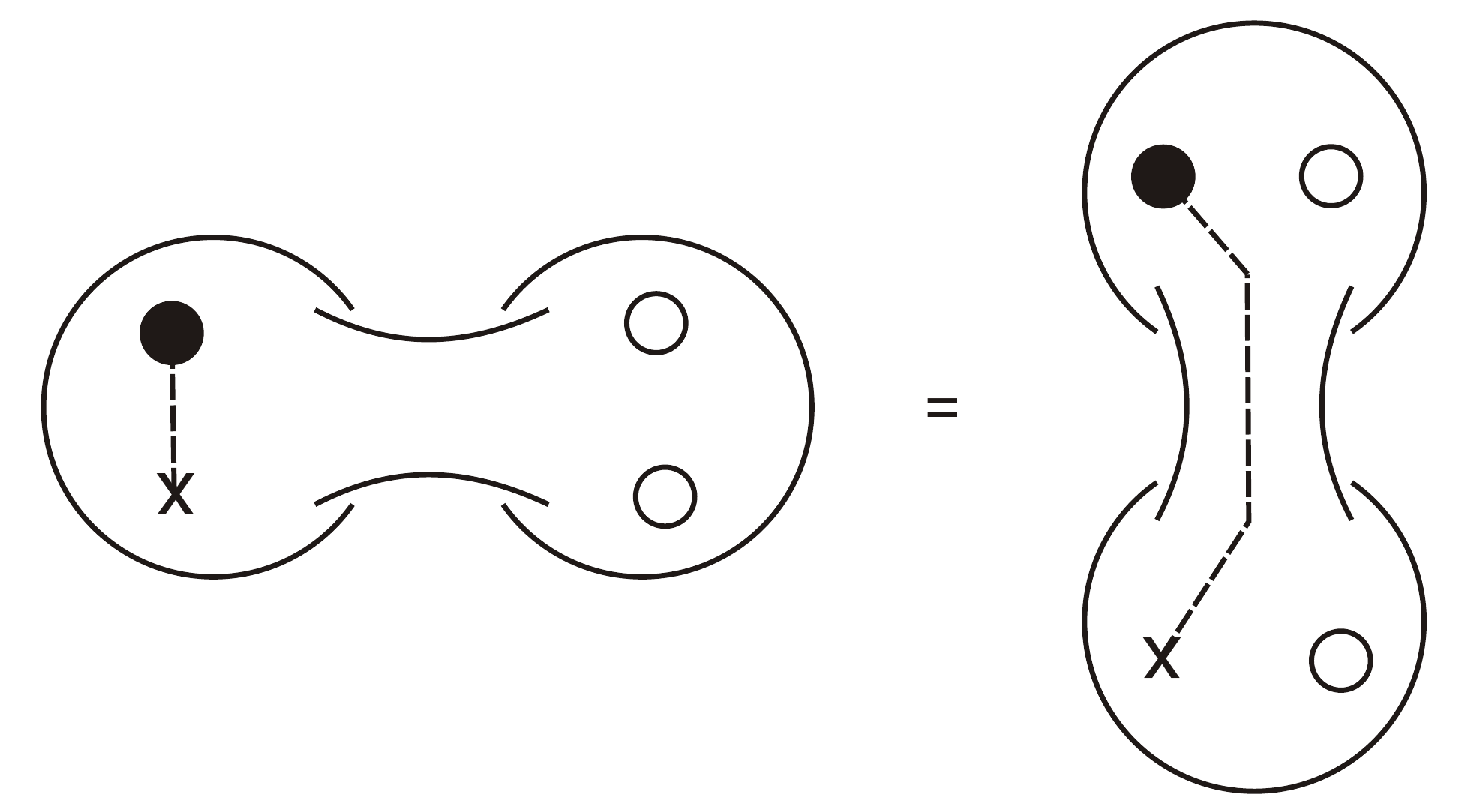}} \quad \quad \quad
\caption{Two examples of S-duality.}\label{Sdualities}
\end{figure}

Gluing two punctures by a cylinder corresponds as always to  gauging the diagonal flavor group.
By gluing the  basic three-punctured spheres introduced so far, see
  figure \ref{basicspheres}, we can generate punctured surfaces of arbitrary topology, subject
  to the  condition that the  $\mathrm{USp}$ punctures must come in an even number,
and be connected pairwise by twist lines.  There can also be twist lines wrapping cycles of the surface.
Different duality frames
correspond to different degeneration
limits of the same surface~\cite{Gaiotto:2009we}.
Basic examples are given in figure~\ref{Sdualities}.

The class of surfaces  generated by the three-punctured spheres  figure \ref{basicspheres} correspond to
 the smallest subclass of theories that contains the alternating linear quivers
with  $\mathrm{SO} (2n)$ and $\mathrm{USp}(2n-2)$ gauge groups and is closed under S-dualities. 
More general superconformal tails will be incorporated
in section  \ref{partclosedpunctures}
by considering more general punctures.

\section{A TQFT with Twist Lines}

In this section we determine the $2d$ TQFT  that computes the $p=0$ limit of the index for the 
 theories described in the previous section. 
Our starting point is the plausible guess that the Macdonald basis should diagonalize the structure constants of the TQFT.
The presence of twist lines and of punctures of both $\mathrm{SO}$ and $\mathrm{USp}$ type
leads however to a more involved structure. 
Indeed we will have to deal with these complications already in the basic example  of free hypers.
Nevertheless, we are led to a natural proposal that passes several checks.

%=====================================%
\subsection{Preliminaries} 
\label{Sec:Index}

%=====================================%
%
Let us recap some basic fact about the ${\cal N}=2$ superconformal index.
We follow the notations and conventions of~\cite{Gadde:2011uv}. The
index~\cite{Kinney:2005ej} is defined
 by a trace over the states on $S^3$ (in the standard radial quantization of the CFT),
\begin{align}\label{index}
\mathcal{I}(p,q,t) = \mathrm{Tr} \, (-1)^F \left(\frac{t}{pq} \right)^r
\left(\frac{p}{q}\right)^{j_1} \left(pq\right)^{j_2} t^R \prod_i
a_i^{f_i} \, .
\end{align}
Here $j_{1,2}$ are the Cartans of the Lorentz group $SU(2)_1 \times
SU(2)_2,$ $r$ is the $U(1)_r$ generator, $R$ is the Cartan of
$SU(2)_R$ and $f_i$ are flavor symmetry charges. The combinations
of superconformal generators in
(\ref{index}) commute with
the supercharge
$\tilde{Q}_{1\dot{-}}$ which carries quantum numbers $j_1=0,$ $j_2 =
-\frac{1}{2},$ $r=-\frac{1}{2}$ and $R=\frac{1}{2}$ and with its conjugate. 
The index 
counts states in the cohomology of
 $\tilde{Q}_{1\dot{-}}$.
 In particular only states satisfying
\begin{align}
2\left\{
\tilde{Q}_{1\dot{-}},\left(\tilde{Q}_{1\dot{-}}\right)^{\dagger}
\right\} = E - 2j_2 - 2R + r = 0
\end{align}
contribute. For theories with a Lagrangian description the index can be
computed as a matrix integral, which can be written schematically as
\begin{align}\label{indexexplicit}
\mathcal{I}(V,p,q,t)=\int[dU] \exp\left( \sum_{n=1}^{+\infty}
\frac{1}{n} \sum_j f^{(j)}\left(p^n,q^n,t^n\right) \cdot
\chi_{\mathcal{R}_j}(U^n,V^n)\right).
\end{align}
Here $U$ collectively denotes all gauge group elements and $V$ all
flavor group elements, while $[dU]$  denotes the invariant
Haar measure for all gauge groups\footnote{In the notation of
appendix \ref{sectionMacdonaldpolyandoperator}, it would be a product
over all gauge groups of factors of the form $[da] \Delta(a).$}. The sum over $j$ runs
over all the $\mathcal{N}=2$ supermultiplets in the Lagrangian. The symbol
$\mathcal{R}_j$ denotes the representation of the $j$-th supermultiplet
under the flavor and gauge groups, and
  $\chi_{\mathcal{R}_j}$ is its
character. Finally $f^{(j)}$ denotes the single-letter indices, which are
given
\begin{align} \label{singleletterpf}
f^V &= -\frac{p}{1-p} - \frac{q}{1-q} + \frac{pq/t -
t}{(1-q)(1-p)},\\
f^{\frac{1}{2}H} &= \frac{\sqrt{t}-pq/\sqrt{t}}{(1-q)(1-p)},
\end{align}
for an $\mathcal{N}=2$ vectormultiplet and an $\mathcal{N}=2$
half-hypermultiplet respectively. 

Several limits of the index (\ref{index}) with enhanced
supersymmetry can be considered. For $p\rightarrow 0,$ one obtains
what was called the Macdonald index in~\cite{Gadde:2011uv}. This
index counts states annihilated by both $\tilde{Q}_{1\dot{-}}$ and
$Q_{1+},$ and thus is a $\frac{1}{4}$ BPS object. If in addition one
takes $q\rightarrow 0,$ one obtains the Hall-Littlewood index, which
counts states annihilated by $\tilde{Q}_{1\dot{-}},$  $Q_{1+}$ and
$Q_{1-}.$ Finally, taking $q=t$ and $p$
arbitrary results in the Schur index, which in fact turns out to be
independent of $p.$ In this paper we will focus on these limits of
enhanced supersymmetry.

%
%=====================================%
\subsection{TQFT structure}
%=====================================%
%

Following the logic of~\cite{Gadde:2009kb}, we wish to identify the TQFT
associated to the index of type $D$ theories of class ${\cal S}$, in the enhanced supersymmetry limit $p=0$
(the ``Macdonald index'').
 In this section we restrict to theories with maximal and empty punctures.
The  basic building blocks are the three-punctured spheres in figure \ref{basicspheres} and the cylinders (propagators).

There are two types of propagators: with or without a twist line running along the cylinder,
corresponding respectively to  $\mathrm{USp}$ and $\mathrm{SO}$ gauge groups.
They are  given by
\bea
\eta_{\mathrm{SO}}(\mathbf{a}_1,\mathbf{a}_2) &=&
\Delta_{\mathrm{SO}}(\mathbf{a}_1) {\cal
I}^V_{\mathrm{SO}}(\mathbf{a}_1)
\delta(\mathbf{a}_1,\mathbf{a}_2^{-1}) \,,\nn\\
\eta_{\mathrm{USp}}(\mathbf{b}_1,\mathbf{b}_2) &=&
\Delta_{\mathrm{USp}}(\mathbf{b}_1) {\cal
I}^V_{\mathrm{USp}}(\mathbf{b}_1)
\delta(\mathbf{b}_1,\mathbf{b}_2^{-1}) \,,
\eea
where $\Delta_{\mathrm{SO}}(\mathbf{a})$ and
$\Delta_{\mathrm{USp}}(\mathbf{a})$ denote the $\mathrm{SO}(2n)$ and
$\mathrm{USp}(2n-2)$ Haar measures respectively. The index for a general
vectormultiplet is known explicitly:
\begin{align}
\mathcal{I}^V &= \mathrm{PE}[f^V \chi^{adj}]\nn \\
&=(q ; q)^{r}_{\infty}(t ; q)^r_{\infty}\prod_{\alpha \in R}(q e^\alpha ; q)_{\infty} (t e^\alpha;q)_{\infty}.
\label{IVroots}
\end{align}
Here the plethystic exponential $\mathrm{PE}$ is defined as
$\mathrm{PE}[f(x_i)]_{x_i} \equiv \exp\left( \sum_{n=1}^{\infty}\frac{1}{n}
f(x_i^n) \right)$, and the $q$-Pochhammer symbol as $(a;q)_{\infty} =
\prod_{j=0}^{\infty}(1-aq^j).$ Usually, we omit the subscript in the plethystic exponential and take
it with respect to all parameters. Moreover,
$f^V$ is the single letter partition function for the
vectormultiplet (\ref{singleletterpf}) in the Macdonald limit
($p\rightarrow 0$),  $\chi^{adj} = r+\sum_{\alpha\in R} e^\alpha$ 
the adjoint character, and $r$ denotes the rank of the group. More
concretely, one has 
for the $\mathrm{SO}(2n)$ and
$\mathrm{USp}(2n-2)$ vectormultiplet indices
\bea
&&\mathcal{I}^V_{\mathrm{SO}}\left(\mathbf{a}\right) =
\left(q,t;q\right)_\infty^{n}\prod_{i < j}\left(q\,a_i^\pm a_j^\pm
,t\,a_i^\pm a_j^\pm ;q\right)_\infty\,,
\label{vecSO}\\
&&\mathcal{I}^V_{\mathrm{USp}}\left(\mathbf{b}\right)=\left(q,t;q\right)_\infty^{n-1}
\prod_{\alpha}\left(q\,b_\alpha^{\pm 2} ,t\,b_\alpha^{\pm 2}
;q\right)_\infty \prod_{\alpha < \beta}\left(q\,b_\alpha^\pm
b_\beta^\pm ,t\,b_\alpha^\pm b_\beta^\pm ;q\right)_\infty \,,
\label{vecUSp}
\eea
where $a_i$, $i=1,\ldots,n$, and $b_\alpha$, $\alpha=1,\ldots,n-1$, denote 
fugacities of $\mathrm{SO}$ and $\mathrm{USp}$ respectively.
Here and throughout the paper we use the condensed notation
that $\pm$ means we take the product over all sign choices and $\left(x_1,\ldots,x_l;q\right)_\infty=\prod_{k=1}^{l}\left(x_k;q\right)_\infty$.

The indices of the different three-punctured spheres are parametrized as in figure
\ref{basicspheres}. All of these are a priori unknown functions of
the flavor symmetry fugacities and $q$ and $t$, except for the free
half-hypermultiplet index $\mathcal{I}\left(\mathbf{a},\mathbf{b}\right)$, which reads
\bea
\mathcal{I}\left(\mathbf{a},\mathbf{b}\right) &=& \mathrm{PE}[f^{\frac{1}{2}H}\chi_{\mathrm{SO}(2n)}^{v}(\mathbf{a})\chi_{\mathrm{USp}(2n-2)}^{f}(\mathbf{b})]\nn\\
&=& \prod_{i=1}^{n}
\prod_{\alpha = 1}^{n-1} \frac{1}{ \left(\sqrt{t}\,a_i^\pm
b_\alpha^\pm ;q \right)_{\infty}} \,. \label{freeH}
\eea
Here $\chi_{\mathrm{SO}(2n)}^{v}$ and $\chi_{\mathrm{USp}(2n-2)}^{f}$ are the
characters of the vector representation of $\mathrm{SO}(2n)$ and the
fundamental representation of $\mathrm{USp}(2n-2)$, respectively.

Following~\cite{Gadde:2011uv}, we expand the free hypermultiplet
index in two complete bases of functions $\lbrace
f^{\lambda}_{\mathrm{SO}}\left(\mathbf{a}\right)\rbrace$ and
$\lbrace
f^{\lambda^{\prime}}_{\mathrm{USp}}\left(\mathbf{b}\right)\rbrace$,
labeled by $\mathrm{SO}(2n)$ and $\mathrm{USp}(2n-2)$
representations respectively,
\beq
\mathcal{I}\left(\mathbf{a},\mathbf{b}\right)=\sum_{\lambda,\lambda^{\prime}}C_{\lambda,\lambda^{\prime}}\,f^{\lambda}_{\mathrm{SO}}\left(\mathbf{a}\right)
f^{\lambda^{\prime}}_{\mathrm{USp}}\left(\mathbf{b}\right)\,.
\label{ExpansionIndexFreeH}
\eeq
These functions are  chosen to be orthonormal under the
measure which appears in each of the propagators
\bea
\oint [d \mathbf{a}_1] \oint [d \mathbf{a}_2] \;
\eta_{\mathrm{SO}} (\mathbf{a}_1,\mathbf{a}_2)\,f^{\lambda_1}_{\mathrm{SO}}({\mathbf a}_1)\,f^{\lambda_2}_{\mathrm{SO}}({\mathbf a}_2)&=&\delta^{\lambda_1 \lambda_2}\,,\nn\\
\oint [d \mathbf{b}_1] \oint [d \mathbf{b}_2] \;
\eta_{\mathrm{USp}} (\mathbf{b}_1,\mathbf{b}_2)\,f^{\lambda_1^\prime}_{\mathrm{USp}}({\mathbf b}_1)\,f^{\lambda_2^\prime}_{\mathrm{USp}}({\mathbf b}_2)&=&\delta^{\lambda_1^\prime \lambda_2^\prime}\,,
\eea
where $[d \mathbf{a}] = \prod_j \frac{da_j}{2 \pi i a_j}$.
We will now show that we can find bases such that the structure
constants, $C_{\lambda,\lambda^{\prime}}$, are ``diagonal'' in the
sense that only some of the $\mathrm{SO}(2n)$ representations appear
in the sum, which are in one-to-one correspondence with all
representations of $\mathrm{USp}(2n-2)$. As in~\cite{Gadde:2011uv}
we consider the following Ans\"{a}tze
\beq
f^{\lambda}_{\mathrm{SO}}({\mathbf a})=\mathcal{K}_{\mathrm{SO}}(\mathbf{a})P^{\lambda}_{\mathrm{SO}}\left(\mathbf{a}\right)\,,\quad\quad
f^{\lambda^\prime}_{\mathrm{USp}}({\mathbf
b})=\mathcal{K}_{\mathrm{USp}}(\mathbf{b})P^{\lambda^{\prime}}_{\mathrm{USp}}\left(\mathbf{b}\right)\,,
\eeq
with the factors $\mathcal{K}$ chosen such that the polynomials $P^{\lambda}$ are orthonormal under the new measures
\bea
&&\hat{ \Delta}_{\mathrm{SO}}({\mathbf a}) \equiv {\mathcal I}^V_{\mathrm{SO}}({\mathbf a})\, {\mathcal K}_{\mathrm{SO}}({\mathbf a}) {\mathcal K}_{\mathrm{SO}}({\mathbf a}^{-1})\Delta_{\mathrm{SO}}({\mathbf a})\,,\\
&&\hat{ \Delta}_{\mathrm{USp}}({\mathbf b}) \equiv {\mathcal I}^V_{\mathrm{USp}}({\mathbf b})\, {\mathcal K}_{\mathrm{USp}}({\mathbf b}) {\mathcal K}_{\mathrm{USp}}({\mathbf b}^{-1})\Delta_{\mathrm{USp}}({\mathbf b}) \,.
\eea
We shall see that if we take the two complete sets of functions
$P^{\lambda}$ \footnote{Note that the Macdonald polynomials as
defined in equation~\eqref{Macdonaldpolygeneral} are only
orthogonal, and so one needs to normalize them to have unit inner
product. $P^{\lambda}$ is used to denote the normalized
polynomials.} to be the Macdonald polynomials for the $\mathrm{SO}$
and $\mathrm{USp}$ groups, the structure constants are
``diagonal''. This means that $\hat{\Delta}$ will be the Macdonald
measure defined in equation~\eqref{MacMeasure}, and this fixes the
$\mathcal{K}$-factors to be
\begin{align}\label{K-factor}
\mathcal{K} = \f{1}{(q;q)_\infty^{r/2}(t;q)_\infty^{r/2}}\prod_{\alpha \in R}\frac{1}{(t
e^\alpha;q)_{\infty}}\,,
\end{align}
where we made use of (\ref{HaarMeasure}) and (\ref{IVroots}) and the obvious invariance under negating roots.

To show the ``diagonality'', we act on the index of the free hyper with
both the $\mathrm{SO}(2n)$ and $\mathrm{USp}(2n-2)$ Macdonald
operators (equations~\eqref{MacOpC} and \eqref{MacOpD}), conjugated
by the respective $\mathcal{K}$-factor, and find that (see appendix
\ref{appendixactionmdoperator})
\beq
\mathcal{K}_{\mathrm{SO}}(\mathbf{a})D_{\mathrm{SO}}\mathcal{K}^{-1}_{\mathrm{SO}}(\mathbf{a}) \mathcal{I}\left(\mathbf{a},\mathbf{b}\right)=
\mathcal{K}_{\mathrm{USp}}(\mathbf{b})D_{\mathrm{USp}}\mathcal{K}^{-1}_{\mathrm{USp}}(\mathbf{b}) \mathcal{I}\left(\mathbf{a},\mathbf{b}\right)\,.
\eeq
Since the Macdonald polynomials are eigenfunctions of $D$, the
functions $f^{\lambda}$ will be eigenfunctions of $\mathcal{K}D
\mathcal{K}^{-1}$ with non-degenerate eigenvalues $c^\lambda$. Using
the expansion of the index this means that
\beq
\sum_{\lambda,\lambda^{\prime}}c^{\mathrm{USp}}_{\lambda^{\prime}}C_{\lambda,\lambda^{\prime}}\,f^{\lambda}_{\mathrm{SO}}\left(\mathbf{a}\right)
f^{\lambda^{\prime}}_{\mathrm{USp}}\left(\mathbf{b}\right)=
\sum_{\lambda,\lambda^{\prime}}c^{\mathrm{SO}}_{\lambda} C_{\lambda,\lambda^{\prime}}\,f^{\lambda}_{\mathrm{SO}}\left(\mathbf{a}\right)
f^{\lambda^{\prime}}_{\mathrm{USp}}\left(\mathbf{b}\right)\,.
\eeq
As noted in appendix~\ref{Sec:ExplicitMac} we have $c^{\mathrm{USp}}_{\left(\lambda_1^\prime,\ldots,\lambda_{n-1}^\prime\right)}=c^{\mathrm{SO}}_{\left(\lambda_1^\prime,\ldots,\lambda_{n-1}^\prime,\lambda_{n-1}^\prime\right)}
$, where $\lambda_i^\prime$ denotes the Dynkin labels of the representation $\lambda^\prime$. This, together with the non-degeneracy of the eigenvalues $c^{\mathrm{SO}}_\lambda$, implies that the expansion of the index in equation~\eqref{ExpansionIndexFreeH} is ``diagonal'' in the sense that for any $\mathrm{USp}(2n-2)$ representation $\lambda^\prime=\left(\lambda_1^\prime,\ldots,\lambda_{n-1}^\prime\right)$
\beq
C_{\lambda,\lambda^\prime} \neq 0\quad \Rightarrow \quad \lambda=\left(\lambda_1^\prime,\ldots,\lambda_{n-1}^\prime,\lambda_{n-1}^\prime\right)\,,
\eeq
hence the sum runs over all $\mathrm{USp}(2n-2)$ representations.

The index of
$T_{\mathrm{SO}(2n)}$ can also be expanded in the set of functions
$\lbrace f^{\lambda}_{\mathrm{SO}}\left(\mathbf{a}\right)\rbrace$
\beq
\label{TSOexp}
\mathcal{I}_{T_{\mathrm{SO}(2n)}}\left(\mathbf{a}_1,\mathbf{a}_2,\mathbf{a}_3\right)=\sum_{\lambda_1,\lambda_2,\lambda_3}B_{\lambda_1,\lambda_2,\lambda_3}\,f^{\lambda_1}_{\mathrm{SO}}\left(\mathbf{a}_1\right)
f^{\lambda_2}_{\mathrm{SO}}\left(\mathbf{a}_2\right)f^{\lambda_3}_{\mathrm{SO}}\left(\mathbf{a}_3\right)\,.
\eeq
Motivated by the structure of the $T_N$ theory index of \cite{Gadde:2011uv,Gaiotto:2012xa} we assume that the basis of functions which diagonalizes the $C_{\lambda,\lambda'}$ also diagonalizes the $B$ structure constants,\footnote{This would follow at once from the approach of \cite{Gaiotto:2012xa}, namely
if one could derive the action of $D$ on the index of  theory ${\cal T}$
 by extracting residues (in flavor fugacities) in the index of a bigger theory ${\cal T}'$. Taking residues in different duality
frames of ${\cal T}'$ one would get the action of  $D$ on different flavor punctures of ${\cal T}$. Invariance of the index under
S-duality would then imply diagonality in the basis of the eigenfunctions of $D$.
As in
 \cite{Gaiotto:2012xa}, we also expect that acting with $D$ on the index
amounts to decorating the $4d$ theory with a BPS surface defect.}

\beq
B_{\lambda_1,\lambda_2,\lambda_3}\neq0 \Rightarrow \lambda_1=\lambda_2=\lambda_3 \,.
\eeq
Note that the sum in \eqref{TSOexp} runs over all $\mathrm{SO}(2n)$ representations.

The S-dualities shown in figure~\ref{Sdualities} can be written in
simple form if we also expand the indices of the three-punctured spheres in
figures~\ref{sphereE} and \ref{sphereA} in the $\lbrace
f^{\lambda}_{\mathrm{SO}}\left(\mathbf{a}\right)\rbrace$ and
$\lbrace
f^{\lambda^{\prime}}_{\mathrm{USp}}\left(\mathbf{b}\right)\rbrace$
bases, and can be used to constrain the structure constants of these
spheres. We write the indices of these spheres as
\bea
\mathcal{I}_{\tilde{T}_{\mathrm{SO}(2n)}}\left(\mathbf{a},\mathbf{b}_1,\mathbf{b}_2\right)&=&
\sum_{\lambda,\lambda'_1,\lambda'_2} A_{\lambda,\lambda'_1,\lambda'_2}
f_{\mathrm{SO}}^\lambda(\mathbf{a})
f^{\lambda_1^{\prime}}_{\mathrm{USp}}\left(\mathbf{b}_1\right)
f^{\lambda_2^{\prime}}_{\mathrm{USp}}\left(\mathbf{b}_2\right)\,,\label{ExpansionA}\\
\mathcal{I}\left(\mathbf{a}\right)&=&\sum_\lambda E_\lambda f_{\mathrm{SO}}^\lambda(\mathbf{a})\,,\label{ExpansionE}
\eea
respectively. The S-duality shown in figure~\ref{Sdual2} written in
terms of the structure constants reads
\beq
\sum_{\lambda_1} C_{\lambda_1, \lambda'_1} B_{\lambda_1,\lambda_2\,,\lambda_3} =\sum_{\lambda'_2} A_{\lambda_2,\lambda'_1,\lambda'_2} C_{\lambda_3,\lambda'_2}\,.
\eeq
Using the ``diagonality'' of $C_{\lambda \lambda'}$ and
$B_{\lambda_1,\lambda_2,\lambda_3}$ this implies that
\beq  C_{\lambda_1 =\lambda'_1 , \lambda'_1} B_{\lambda_1 =
\lambda'_1,\lambda_2\,,\lambda_3} \hat{\delta}_{\lambda_2,
\lambda'_1 } \hat{\delta}_{\lambda_3, \lambda'_1 }=
A_{\lambda_2,\lambda'_1,\lambda'_3} C_{\lambda_3 =
\lambda'_3,\lambda'_3} \hat{\delta}_{\lambda_3, \lambda'_3 }\,. \eeq

Here we used the shorthand notation that $\lambda=\lambda'$ means
$\lambda=\left(\lambda_1^\prime,\ldots,\lambda_{n-1}^\prime,\lambda_{n-1}^\prime\right)$
and $\hat{\delta}_{\lambda, \lambda'}$ imposes the same condition.
Now we find that $A_{\lambda_2,\lambda'_1,\lambda'_3}$ vanishes
unless $\lambda'_1=\lambda'_3$ and $\lambda_2=\lambda'_1$, This
relation also fixes the value of the $A$ structure constants as
\beq A_{\lambda,
\lambda'_1,\lambda'_2}=\hat{\delta}_{\lambda,\lambda'_1}
\delta_{\lambda'_1,\lambda'_2}
B_{\lambda_1=\lambda'_1,\lambda_2=\lambda'_1,\lambda_3=\lambda'_1}
\,. \label{SdualAstr}\eeq
The ``diagonality'' of the $A$ structure constants means that the sum in equation~\eqref{ExpansionA} runs only over $\mathrm{USp}(2n-2)$ representations.

A similar reasoning will allow us to obtain the $E$ structure constants from the S-duality in figure~\ref{Sdual1}, which reads
\beq
\sum_{\lambda'} C_{\lambda_2, \lambda'} C_{\lambda_3, \lambda'} =\sum_{\lambda_1} B_{\lambda_1,\lambda_2,\lambda_3} E_{\lambda_1}\,.
\eeq
The ``diagonality'' of $C$ and $B$ implies that $ E_{\lambda}$ vanishes unless it is of the form $\lambda=\left(\lambda_1,\ldots,\lambda_{n-1},\lambda_{n-1}\right)$, and so the sum in equation~\eqref{ExpansionE} does not run over all $\mathrm{SO}(2n)$ representations. The value of $E$ is also fixed by this duality to be
\beq
E_{\lambda}=\hat{\delta}_{\lambda=\lambda'} \f{C_{\lambda=\lambda',\lambda'}^2}{B_{\lambda_1=\lambda',\lambda_2=\lambda',\lambda_3=\lambda'}}\,.
\label{SdualEstr}
\eeq

Having introduced the structure constants for the different basic
spheres, we now turn to evaluating them in some special cases. In our calculations we have
made extensive use of the LieART Mathematica package of~\cite{2012arXiv1206.6379F}, and use the determinantal formula of~\cite{vanDiejen} to compute the Macdonald polynomials.
%=====================================%
\subsection{$D_2$ theories}
%=====================================%
Although not an honest member of the $D$-series, we will start our
discussion of the structure constants with $D_2 \equiv A_1 \times A_1.$ Due to the low
rank, it is possible to obtain exact results in the Hall-Littlewood and Schur limits.
These theories also allow for an interesting interpretation in terms of
$A_1$ theories.
%=====================================%
\subsubsection{Hall-Littlewood limit}
%=====================================%
The simplest limit of enhanced supersymmetry is the Hall-Littlewood
limit, {\it i.e.} $p\to0\,, q\to0,$ since then the $q$-Pochhammer symbols in
equations~\eqref{vecSO}-\eqref{freeH} simplify, as $(x;0)_\infty=1-x$.
Moreover, all sums over representations will be geometric
progressions.

Without much ado, we write the index of the free hypermultiplet for
the $D_2$ theory as
\begin{align}
&\mathcal{I}\left(\mathbf{a},b\right)=\nn\\
&\mathcal{A}(\tau)\sum_{\lambda'=0}^{\infty}
\f{\mathcal{K}_{\mathrm{USp}}(\mathbf{\times})
P^{(\lambda')}_{HL\,\mathrm{USp}}\left(\tau\mid\tau\right)\,\mathcal{K}_{\mathrm{USp}}(b)P^{(\lambda')}_{HL\,\mathrm{USp}}\left(b\mid\tau\right)\mathcal{K}_{\mathrm{SO}}(\mathbf{a})P^{(\lambda',\,\lambda')}_{HL\,\mathrm{SO}}\left(\mathbf{a}\mid\tau\right)}{P^{(\lambda',\,\lambda')}_{HL\,\mathrm{SO}}\left(1,\tau^2\mid\tau\right)}
\,,
\end{align}
where we defined
\beq
\mathcal{K}_{\mathrm{USp}}(\mathbf{\times})= \f{\sqrt{1-\tau^2}}{1-\tau^4}\,,\quad \quad \mathcal{A}(\tau)=\f{\left(1-\tau^4\right)^2}{1-\tau^2}\,.
\eeq
The equality of the above expression with equation~\eqref{freeH} for
$q=0$ was checked by performing the sum over representations, which
is geometric. We can read off the structure constant $C_{\lambda=\lambda'\,,\lambda'}$ to be
\beq \label{Cstructconstant}
C_{\lambda=\lambda'\,,\lambda'}=\f{\mathcal{A}(\tau)
\mathcal{K}_{\mathrm{USp}}(\mathbf{\times})P^{(\lambda')}_{HL\,\mathrm{USp}}\left(\tau\mid\tau\right)}{
\mathrm{dim}^{\mathrm{SO}}_{\tau^2}(\lambda = \lambda')}\,,
\eeq
where we used the Hall-Littlewood limit of the $\mathrm{SO}(4)$ $(q,t)$-dimension formula. 

In general, the $(q,t)$-dimension (or Macdonald dimension) for $\mathrm{SO}(2n)$ is given by
\begin{align}
\mathrm{dim}^{\mathrm{SO}}_{q,t}(\lambda) = P^{\lambda}_{M\,\mathrm{SO}}\left(1,t,\ldots,t^{n-1}\mid
q,t\right).
\end{align}
The usual limits apply: $q\rightarrow 0$ gives the Hall-Littlewood or $t$-dimension $\mathrm{dim}^{\mathrm{SO}}_{t}$ and $q=t$ gives the Schur dimension $\mathrm{dim}^{\mathrm{SO}}_{q}$, also known as the $q$-dimension. In the Hall-Littlewood limit we write $t=\tau^2.$ 

The expression (\ref{Cstructconstant}) for the structure constant, as well as the ones below for the other spheres, will hold in the other limits and for higher rank if naturally modified and generalized.

Since $\mathrm{USp}(2) = \mathrm{SU}(2),$ we can apply the procedure
of~\cite{Gaiotto:2012xa} to close the $\mathrm{USp}(2)$ puncture and
obtain the index of the sphere with one $\mathrm{SO}(4)$ puncture
and two closed $\mathrm{USp}$ punctures. This index is computed as
$2\, \mathcal{I}_V \mathrm{Res}_{b=\tau}\left( \f{1}{b}
\mathcal{I}\left(\mathbf{a},b\right)\right)$, with
$\mathcal{I}_V=\mathrm{\mathrm{PE}}\left[ f_V\right].$ We find
\beq
\mathcal{I}(\mathbf{a})=\mathcal{A}(\tau)\sum_{\lambda'=0}^{\infty}
\f{\left(\mathcal{K}_{\mathrm{USp}}(\mathbf{\times})P^{(\lambda')}_{HL\,\mathrm{USp}}\left(\tau\mid\tau\right)\right)^2\mathcal{K}_{\mathrm{SO}}(\mathbf{a})P^{(\lambda',\,\lambda')}_{HL\,\mathrm{SO}}\left(\mathbf{a}\mid\tau\right)}{P^{(\lambda',\,\lambda')}_{HL\,\mathrm{SO}}\left(1,\tau^2\mid\tau\right)}
\,.
\eeq
Notice that the index for this sphere vanishes when
summed over representations. Since it has
$n_V=-3$ and $n_H=0$ it only makes sense as part of a larger theory,
{\it e.g.} in figure~\ref{Sdual1}. The $E_{\lambda}$ structure constant is
\beq
E_{\lambda=\lambda'}=\f{\mathcal{A}(\tau)
\mathcal{K}_{\mathrm{USp}}(\mathbf{\times})P^{(\lambda')}_{HL\,\mathrm{USp}}\left(\tau\mid\tau\right) \mathcal{K}_{\mathrm{USp}}(\mathbf{\times})P^{(\lambda')}_{HL\,\mathrm{USp}}\left(\tau\mid\tau\right) }{
\mathrm{dim}^{\mathrm{SO}}_{\tau^2}(\lambda = \lambda')}\,.
\eeq

Considering the relations imposed by S-duality in~\eqref{SdualEstr}
and~\eqref{SdualAstr} one can now also write down the index for the
$T_{\mathrm{SO}(4)}$ theory and the $\tilde{T}_{\mathrm{SO}(4)}$
theory, respectively, as
\beq
\mathcal{I}_{T_{\mathrm{SO}(4)}}\left(\mathbf{a}_1,\mathbf{a}_2,\mathbf{a}_3\right)=\mathcal{A}(\tau)
\sum_{\lambda_1,\,\lambda_2=0}^{\infty}\f{1}{P^{(\lambda_1,\,\lambda_2)}_{HL\,\mathrm{SO}}\left(1,\tau^2\mid\tau\right)}\prod_{i=1}^3
\,\mathcal{K}_{\mathrm{SO}}(\mathbf{a}_i)P^{(\lambda_1,\,\lambda_2)}_{HL\,\mathrm{SO}}\left(\mathbf{a}_i\mid\tau\right)\,
\eeq
and
\beq
\mathcal{I}_{\tilde{T}_{\mathrm{SO}(4)}}\left(\mathbf{a},b_1,b_2\right)=
\mathcal{A}(\tau)\sum_{\lambda'=0}^{\infty}
\f{\mathcal{K}_{\mathrm{SO}}(\mathbf{a})P^{(\lambda',\,\lambda')}_{HL\,\mathrm{SO}}\left(\mathbf{a}\mid\tau\right)\,\prod_{i=1}^{2}\mathcal{K}_{\mathrm{USp}}(b_i) P^{(\lambda')}_{HL\,\mathrm{USp}}\left(b_i\mid\tau\right)}{P^{(\lambda',\,\lambda')}_{HL\,\mathrm{SO}}\left(1,\tau^2\mid\tau\right)}
\,.
\eeq

These results were also checked against the construction in
figure~\ref{TSO} and \ref{TSOtilde}. The index for the linear quivers
was expanded in powers of $\tau$ and compared to the expansion of
the $T_{\mathrm{SO}(4)}$ (respectively $\tilde{T}_{\mathrm{SO}(4)}$)
theory glued to its three tails. On the other hand, the independent
construction of the $T_{\mathrm{SO}(4)}$ and
$\tilde{T}_{\mathrm{SO}(4)}$ provides evidence of the S-duality in
figure~\ref{Sdual2}. The structure constants for these theories are
\beq
B_{\lambda,\lambda\,,\lambda}=\f{\mathcal{A}(\tau)}{
\mathrm{dim}^{\mathrm{SO}}_{\tau^2}(\lambda)}\,,\qquad
A_{\lambda=\lambda'\,,\lambda'\,,\lambda'}=\f{\mathcal{A}(\tau)}{
\mathrm{dim}^{\mathrm{SO}}_{\tau^2}(\lambda = \lambda')}\,.
\eeq
For the $T_{\mathrm{SO}(4)}$ theory, we recall that, from equation~\eqref{nVTSO},
the effective number of
hypermultiplets in this theory is 8, and the effective number of
vectormultiplets is 0. In fact, we claim that it corresponds to the index of one half-hypermultiplet in the
$\mathbf{2}\times\mathbf{2}\times\mathbf{2}$ and another in the
$\mathbf{\bar{2}}\times\mathbf{\bar{2}}\times\mathbf{\bar{2}}$ of
the flavor symmetry group $\mathrm{SO}(4)^3$. This was checked in a $\tau$ expansion, and numerically after performing the sum over representations. We will discuss this further in Sec.~\ref{D2asA1}.

%=====================================%
\subsubsection{Schur limit}
%=====================================%
The index for the free half-hypermultiplet in the Schur limit is
given by
\beq
\mathcal{I}\left(\mathbf{a},b\right)=\mathcal{A}(q)
\sum_{\lambda'=0}^{\infty}\f{\mathcal{K}_{\mathrm{USp}}(\mathbf{\times})\chi^{(\lambda')}_{\mathrm{USp}}\left(q^{1/2}\right)\mathcal{K}_{\mathrm{USp}}(b)\chi^{(\lambda')}_{\mathrm{USp}}\left(b\right)\mathcal{K}_{\mathrm{SO}}(\mathbf{a})\chi^{(\lambda',\,\lambda')}_{\mathrm{SO}}\left(\mathbf{a}\right)}{\chi^{(\lambda',\,\lambda')}_{\mathrm{SO}}\left(1,q\right)}\,,
\eeq
where we define
\beq
\mathcal{K}_{\mathrm{USp}}(\mathbf{\times})=\f{1}{(q^2;q)}\,,\quad \quad \mathcal{A}(q)=(q^2;q)^2_\infty\,.
\eeq
The equality of the above expression with equation~\eqref{freeH} for
$q=t$ was proved exactly by comparing the analytic structure of both
sides. In fact, as we will see later in Sec.~\ref{D2asA1}, a
redefinition of the fugacities maps this theory onto the $T_2$
theory discussed in \cite{Gadde:2011uv}, and therefore the proof is
immediate from the similar proof there.

The index of the $T_{\mathrm{SO}(4)}$ theory is
\beq
\mathcal{I}_{T_{\mathrm{SO}(4)}}\left(\mathbf{a}_1,\mathbf{a}_2,\mathbf{a}_3\right)= \mathcal{A}(q)
\sum_{\lambda_1,\,\lambda_2=0}^{\infty}\f{1}{\chi^{(\lambda_1,\,\lambda_2)}_{\mathrm{SO}}\left(1,q\right)}\prod_{i=1}^3
\,\mathcal{K}_{\mathrm{SO}}(\mathbf{a}_i)\chi^{(\lambda_1,\,\lambda_2)}_{\mathrm{SO}}\left(\mathbf{a}_i\right)\,.
\eeq
Again we checked this expression using the construction of
figure~\ref{TSO} by expanding the index for both sides in $q$.
S-duality then fixes the index for the other three-punctured spheres.
The index for $\tilde{T}_{\mathrm{SO}(4)}$ was independently checked
against the construction of figure~\ref{TSOtilde} in a $q$ expansion.
%
%=====================================%
\subsubsection{Macdonald limit}
%=====================================%
%
Let us finally state our claim in the Macdonald limit. For the free half-hypermultiplet we have
\small
\bea
&&\mathcal{I}\left(\mathbf{a},b\right)=\nn\\
&&\mathcal{A}(q,t)\sum_{\lambda'}\f{
\mathcal{K}_{\mathrm{USp}}(\mathbf{\times})P^{(\lambda')}_{M\,\mathrm{USp}}\left(t^{1/2}\mid q,t\right)
\mathcal{K}_{\mathrm{USp}}(b)P^{(\lambda')}_{M\,\mathrm{USp}}\left(b\mid q,t\right)
\mathcal{K}_{\mathrm{SO}}(\mathbf{a})P^{(\lambda',\lambda')}_{M\,\mathrm{SO}}\left(\mathbf{a}\mid q,t\right)}
{P^{(\lambda',\lambda')}_{M\,\mathrm{SO}}\left(1,t\mid q,t\right)}
\,,\nn\\
&&\mathcal{A}(q,t)=\left(t^2;q\right)_\infty^2
\left(\f{ (q;q)_\infty}{(t;q)_\infty}\right) \,,\qquad
\mathcal{K}_{\mathrm{USp}}(\mathbf{\times})=\left(\f{ (t;q)_\infty}{(q;q)_\infty}\right)^{\f{1}{2}}
\left(t^{2};q\right)_\infty^{-1}\,.
\eea
\normalsize
Some of the structure constants were computed in an expansion in $q$ and $t$, and the equality was also checked numerically, truncating the sum at high enough $\lambda'$.

For the $T_{\mathrm{SO}(4)}$ theory we have
\beq \label{TSO4}
\mathcal{I}_{T_{\mathrm{SO}(4)}}\left(\mathbf{a}_1,\mathbf{a}_2,\mathbf{a}_3\right)=\sum_{\lambda_1,\lambda_2}\f{\mathcal{A}(q,t)}{P^{(\lambda_1,\lambda_2)}_{M\,\mathrm{SO}}\left(1,t\mid q,t\right)}
\prod_{i=1}^3
\,\mathcal{K}_{\mathrm{SO}}(\mathbf{a}_i)P^{(\lambda_1,\lambda_2)}_{M\,\mathrm{SO}}\left(\mathbf{a}_i\mid q,t\right)\,,\nn\\
\eeq
where the sum runs over all ${\mathrm{SO}(4)}$ representations.
%
%=====================================%
\subsubsection{$D_2$ theories in terms of $A_1$ theories } \label{D2asA1}
%=====================================%
%
Since locally $\mathrm{SO}(4) = \mathrm{SU}(2)\times \mathrm{SU}(2),$ $\mathrm{USp}(2) = \mathrm{SU}(2)$ and in the superconformal tails $\mathrm{SO}(3) = \mathrm{SU}(2),$  one expects that the $D_2$ theories can also be described in terms of the $A_1$ theory \cite{Tachikawa:2009rb}. 

As mentioned before, one naturally considers alternating quivers (see also section \ref{classification}), which for $D_2$ theories have gauge groups
\begin{align}
\mathrm{SO}(3)\times \mathrm{USp}(2)\times \mathrm{SO}(4)\times \ldots \times \mathrm{USp}(2)\times \mathrm{SO}(3).
\end{align}
Calling the number of $\mathrm{SO}(4)$ gauge groups $s,$ the corresponding Riemann surface is a sphere with $2s+6$ empty $\mathrm{USp}(2)$ punctures $\mathbf{\times}$ connected in pairs by $s+3$ twist lines. In $A_1$ language, it corresponds to a genus $s+2$ Riemann surface, which however, since the couplings of the two $\mathrm{SU}(2)$s within $\mathrm{SO}(4)$ are not independent, is hyperelliptic \cite{Tachikawa:2009rb}. 

In writing $\mathrm{SO}(4)$ as the product of two $\mathrm{SU}(2)$s, the Macdonald polynomials decompose as
\begin{align}\label{factorize}
P_{M\, \mathrm{SO}}^{(\lambda_1,\lambda_2)}(a_1,a_2 \mid q,t) = P_{M\, \mathrm{SU}}^{(\lambda_1)}\left(\sqrt{\frac{a_1}{a_2}},\sqrt{\frac{a_2}{a_1}} \mid q,t\right)P_{M\, \mathrm{SU}}^{(\lambda_2)}\left(\sqrt{a_1a_2},\frac{1}{\sqrt{a_1a_2}} \mid q,t\right).
\end{align}
As a consequence, an $\mathrm{SO}(4)$ puncture splits into two $\mathrm{SU}(2)$ punctures, and the $\mathrm{SO}(4)$ propagator turns into two $\mathrm{SU}(2)$ propagators connecting the two $\mathrm{SU}(2)$ punctures separately.

Let us now rewrite the index for the basic $D_2$ spheres in terms of $A_1$ quantities. First, we have that the index of the free $D_2$ half-hypermultiplet $\mathcal{I}_{D_2}$ trivially equals the index of the $T_2$ theory $\mathcal{I}_{T_2}$ if the fugacities are appropriately identified:
\begin{align}
\mathcal{I}_{D_2}((a_1,a_2),b) = \mathcal{I}_{T_2}\left(\sqrt{a_1
a_2},\sqrt{\frac{a_1}{a_2}},b\right).
\end{align}
Second, the index for the $T_{\mathrm{SO}(4)}$ theory is, as was mentioned already, equal to the index of one half-hypermultiplet in the $\mathbf{2}\times\mathbf{2}\times\mathbf{2}$ and another in the $\mathbf{\bar{2}}\times\mathbf{\bar{2}}\times\mathbf{\bar{2}}$ of the flavor symmetry group $\mathrm{SO}(4)^3.$ This can be rewritten in terms of a product of indices of the $T_2$ theory.
\begin{align}
\label{result}
&\mathcal{I}_{T_{\mathrm{SO}(4)}}((a_1,a_2),(A_1,A_2),(\alpha_1,\alpha_2))
=\nn\\
&\mathcal{I}_{T_2}\left(\sqrt{a_1 a_2},\sqrt{A_1
A_2},\sqrt{\alpha_1
\alpha_2}\right)\mathcal{I}_{T_2}\left(\sqrt{\frac{a_1}{a_2}},\sqrt{\frac{A_1}{A_2}},\sqrt{\frac{\alpha_1}{\alpha_2}}\right).
\end{align}
This factorized expression can also be easily checked by applying \eqref{factorize} to the expression for the index in \eqref{TSO4} and noticing that the $\mathcal{K}$-factors and $\mathcal{A}$ decompose nicely in the product of the corresponding $\mathrm{SU}(2)$ quantities. 
The result (\ref{result}) can be understood from the construction of the $T_{\mathrm{SO}(4)}$ theory in figure \ref{TSO}. Here we have the three-punctured sphere with three maximal $\mathrm{SO}(4)$ punctures, of which from each the diagonal $\mathrm{SO}(3)$ subgroup is gauged. The curve corresponding to the linear quiver describing this theory is a sphere with six empty $\mathrm{USp}$ punctures $\mathbf{\times}.$ The corresponding $A_1$ hyperelliptic curve is a genus two surface, which can be thought of as gluing together two three-punctured spheres. One can now observe that each of these three-punctured spheres contains one of the two $\mathrm{SU}(2)$ punctures in which each of the three $\mathrm{SO}(4)$ punctures split and gauging the diagonal $\mathrm{SO}(3)$ subgroup is indeed obtained by gluing these together.

Third, the index for the $\tilde{T}_{\mathrm{SO}(4)}$ theory is equal to the $A_1$ index of the four-punctured sphere if again the $\mathrm{SO}(4)$ fugacities are appropriately identified,
\begin{align}
\label{otherresult}
&\mathcal{I}_{\tilde{T}_{\mathrm{SO}(4)}}((a_1,a_2),b,c) = \mathcal{I}_{\text{4-punctured}}^{\mathrm{SU}(2)} \left(\sqrt{a_1 a_2} ,\sqrt{\frac{a_1}{a_2}} , b,c \right).
\end{align}
Notice that the number of effective hyper and vectormultiplets is the same for both theories, as can be inferred from \eqref{nvhtilde}, namely eight and three respectively. 
We can also understand the result (\ref{otherresult}) from the construction in figure \ref{TSOtilde}. There we have the three-punctured sphere with one maximal $\mathrm{SO}(4)$ puncture, of which the diagonal $\mathrm{SO}(3)$ subgroup is gauged, and two maximal $\mathrm{USp}(2)$ punctures, which each are connected to an $\mathrm{SO}(3)$ gauge group. The linear quiver describing this theory corresponds to a sphere with eight empty $\mathrm{USp}$ punctures $\mathbf{\times},$ and the corresponding $A_1$ hyperelliptic curve is a genus three surface. This surface can be thought of as originating from a four-punctured sphere of which two punctures are glued to each other and the two other punctures are glued to a torus. As before, gluing the two punctures to each other corresponds to gauging the diagonal $\mathrm{SO}(3)$ subgroup, and gluing a torus corresponds to connecting to an $\mathrm{SO}(3)$ gauge group (since the adjoint of $\mathrm{SO}(3)$ equals the adjoint of $\mathrm{SU}(2)$).

Finally, one can also interpret the partially closed colored box-shaped $\mathrm{SO}(4)$ puncture in $A_1$ language. We will do so in section \ref{kfact}.

%
%=====================================%
\subsection{$D_3$ theories}
%=====================================%
%
We claim that in the $D_3$ case the index of the free hypermultiplet is given by
\small
\bea
&&\mathcal{I}\left(\mathbf{a},\mathbf{b}\right)=\nn\\
&&\mathcal{A}(q,t)\sum_{\lambda}\f{
\mathcal{K}_{\mathrm{USp}}(\mathbf{\times})P^{\lambda'}_{M\,\mathrm{USp}}\left(t^{1/2},t^{3/2}| q,t\right)
\mathcal{K}_{\mathrm{USp}}(\mathbf{b})P^{\lambda'}_{M\,\mathrm{USp}}\left(\mathbf{b}\mid q,t\right)
\mathcal{K}_{\mathrm{SO}}(\mathbf{a})P^{\lambda=\lambda'}_{M\,\mathrm{SO}}\left(\mathbf{a}\mid q,t\right)}
{P^{\lambda=\lambda'}_{M\,\mathrm{SO}}\left(1,t,t^2 \mid q,t\right)}
\,,\nn\\
&&\mathcal{A}(q,t)=\left(t^3;q\right)_\infty
\left(\f{ (q;q)_\infty}{(t;q)_\infty}\right)^{3/2} \prod_{l=1}^{2}
\left(t^{2l};q\right)_\infty\,,\nn\\
&&\mathcal{K}_{\mathrm{USp}}(\mathbf{\times})=\f{ (t;q)_\infty}{(q;q)_\infty} \prod_{l=1}^{2}
\left(t^{2l};q\right)_\infty^{-1}\,.
\eea
\normalsize
In the Hall-Littlewood limit this equality was checked numerically after performing the sum exactly. In the Schur limit these coefficients were checked in a $q$ expansion for several representations, and the equality of the index in this limit was also checked numerically, truncating the sum over representations at a high enough $\lambda'$.

Since $\mathrm{SO}(6)=\mathrm{SU}(4)$, the $T_{\mathrm{SO}(6)}$ theory is the same as the $T_4$ theory (the sphere with 3 maximal $\mathrm{SU}(4)$ punctures), the index of which was computed in~\cite{Gadde:2011uv}. The index for this theory can then just be found from equation~(7.11) of ~\cite{Gadde:2011uv} by re-writing the $\mathrm{SU}(4)$ Macdonald polynomials as $\mathrm{SO}(6)$ polynomials. Denoting the $\mathrm{SU}(4)$ fugacities by $(c_1, c_2, c_3, c_4)$, subject to the constraint $c_1 c_2 c_3 c_4=1$, we have $P^{\tilde{\lambda}}_{M\,\mathrm{SU}}(c_1,c_2,c_3,c_4)=P^{\lambda}_{M\,\mathrm{SO}}(c_1 c_2,c_1 c_3,c_2 c_3)$, where $\tilde{\lambda}$ denotes the $\mathrm{SU}(4)$ representation corresponding to the $\lambda$ $\mathrm{SO}(6)$ representation.
The index then becomes, with the sum running over all ${\mathrm{SO}(6)}$ representations,
\beq
\mathcal{I}_{T_{\mathrm{SO}(6)}}\left(\mathbf{a}_1,\mathbf{a}_2,\mathbf{a}_3\right)=\sum_{\lambda}\f{\mathcal{A}(q,t)}{P^{\lambda}_{M\,\mathrm{SO}}\left(1,t,t^2
\mid q,t\right)}
\prod_{i=1}^3
\,\mathcal{K}_{\mathrm{SO}}(\mathbf{a}_i)P^{\lambda}_{M\,\mathrm{SO}}\left(\mathbf{a}_i\mid q,t\right)\,.\nn\\
\eeq
S-duality then fixes the index of the $\tilde{T}_{\mathrm{SO}(6)}$,
and as in the $D_2$ case this index was independently checked using
the construction of figure~\ref{TSOtilde} both in the
Hall-Littlewood and Schur limits (in a $\tau$ and $q$ expansion
respectively).
%
%=====================================%
\subsection{$D_n$ theories}\label{Dntheories}
%=====================================%
%
Let us finally state our conjecture for the $D_n$ theory in the Macdonald limit\footnote{Note that the $\mathbb{Z}_3$ outer-automorphism of $D_4$ allows for a richer structure, see \cite{Tachikawa:2010vg}. We will not consider this $\mathbb{Z}_3$ twist here.}. The index associated to the free hypermultiplet in the bifundamental of $\mathrm{SO}(2n)\times \mathrm{USp}(2n-2)$ is
\bea
\mathcal{I}\left(\mathbf{a},\mathbf{b}\right)&=&\sum_{\lambda}\f{\mathcal{A}(q,t)}{P^{\lambda=\lambda'}_{M\,\mathrm{SO}}\left(1,t,\ldots,t^{n-1} \mid q,t\right)}\mathcal{K}_{\mathrm{USp}}(\mathbf{\times})P^{\lambda'}_{M\,\mathrm{USp}}\left(t^{1/2},t^{3/2},\ldots,t^{n-3/2}\mid q,t\right)\times\nn\\
&&\times\mathcal{K}_{\mathrm{USp}}(\mathbf{b})P^{\lambda'}_{M\,\mathrm{USp}}\left(\mathbf{b}\mid q,t\right)
\mathcal{K}_{\mathrm{SO}}(\mathbf{a})P^{\lambda=\lambda'}_{M\,\mathrm{SO}}\left(\mathbf{a}\mid q,t\right)
\,,\nn\\
\mathcal{A}(q,t)&=&\left(t^n;q\right)_\infty
\left(\f{ (q;q)_\infty}{(t;q)_\infty}\right)^{n/2} \prod_{l=1}^{n-1}
\left(t^{2l};q\right)_\infty\,,\nn\\
\mathcal{K}_{\mathrm{USp}}(\mathbf{\times})&=&\left(\f{ (t;q)_\infty}{(q;q)_\infty}\right)^{\f{n-1}{2}} \prod_{l=1}^{n-1}
\left(t^{2l};q\right)_\infty^{-1}\label{KfactorminimalUSP}\,.
\eea
This expression was checked numerically for $D_4$ in the
Hall-Littlewood and Schur limits by truncating the sum over
representations. For the $T_{\mathrm{SO}(2n)}$ theory
we conjecture 
\beq
\mathcal{I}_{T_{\mathrm{SO}(2n)}}\left(\mathbf{a}_1,\mathbf{a}_2,\mathbf{a}_3\right)=\sum_{\lambda}\f{\mathcal{A}(q,t)}{P^{\lambda}_{M\,\mathrm{SO}}\left(1,t,\ldots,t^{n-1}
\mid q,t\right)}
\prod_{i=1}^3
\,\mathcal{K}_{\mathrm{SO}}(\mathbf{a}_i)P^{\lambda}_{M\,\mathrm{SO}}\left(\mathbf{a}_i\mid q,t\right)\,,\nn\\
\eeq
where the sum runs over all ${\mathrm{SO}(2n)}$ representations. The
genus $g$ partition function for the pure ${\mathrm{SO}(2n)}$ theory
is
\begin{align}
\mathcal{I}_{T_{\mathrm{SO}(2n)}}&=\sum_{\lambda} \left(\frac{\mathcal{A}(q,t)}{\mathrm{dim}^{SO}_{q,t}(\lambda)}\right)^{2g-2}\\
&=\left((q;q)_\infty (t;q)_\infty\right)^{n(g-1)}
\left(\f{(t;q)_\infty}{(t^n;q)_\infty}\prod_{l=1}^{n-1}
\f{\left(t;q\right)_\infty}{\left(t^{2l};q\right)_\infty}\right)^{2-2g}
\sum_\lambda
\f{1}{\left(\mathrm{dim}^{SO}_{q,t}(\lambda)\right)^{2g-2}}\nn\,.
\end{align}

As in~\cite{Gadde:2011uv}, for the slice $t=q^\beta$ this result
appears to be related to the partition function $Z(\mathcal{C} \times
S^1)$ of a refinement of level $k$ Chern-Simons theory discussed
in~\cite{Aganagic:2012au}. For a genus $g$ Riemann surface, $\mathcal{C}$, $Z$ is given by
\begin{align}
Z(\mathcal{C} \times S^1) = \sum_{\lambda} \frac{1}{\left(
S_{0\lambda}\right)^{2g-2}},
\end{align}
where $S_{0\lambda}$ is given by
\begin{align}
S_{0\lambda} = S_{00}\, \mathrm{dim}^{\mathrm{SO}}_{q,t}(\lambda),
\end{align}
and $S_{00}$ by\footnote{Following \cite{Gadde:2011uv} we only kept the interesting $q$ and $t$ dependent part of $S_{00}$ (see also equation (7.5) of \cite{Aganagic:2012au}). The full three-sphere partition function is
\begin{align}
S_{00} = i^{|\Delta_+|} |P/Q|^{-1/2} (k + \beta
y)^{-n/2} q^{-\beta(\beta-1) |\Delta_+|/4} t^{-\beta \left\langle \rho , \rho\right\rangle} \prod_{m=0}^{\beta -1} \prod_{\alpha
> 0} \left( 1
- q^{m}t^{\left \langle \alpha,\rho \right \rangle}\right)\,,
\label{S002}
\end{align}
where $|\Delta_+|$ is the number of positive roots, $y$ is the dual
coxeter number and $|P/Q|$ denotes the number of elements in the
fundamental cell of the quotient lattice $P/Q.$ Note that with \eqref{S002} there is an extra overall factor when comparing the partition function on $\mathcal{C}\times S^1$ with the superconformal index for the Riemann surface $\mathcal{C}.$ This is precisely the same factor that appears when comparing the $A$-type index of \cite{Gadde:2011uv} with the complete refined Chern-Simons partition function on $\mathcal{C}\times S^1$ for $\mathrm{SU}(N)$.}
\begin{align}
S_{00} = \prod_{m=0}^{\beta -1} \prod_{\alpha
> 0} \left( 1- q^{m}t^{\left \langle \alpha,\rho \right \rangle}\right).\label{S00}
\end{align}
The parameters $q,t$
are identified as $q = \exp\left(\frac{2\pi i}{k + \beta y}\right)$
and $t = \exp\left(\frac{2\pi i \beta}{k + \beta y}\right).$ Also
note that we consider the Macdonald polynomials to be normalized,
and so $\mathrm{dim}^{\mathrm{SO}}_{q,t}(\lambda)$ here is
$\mathrm{dim}^{\mathrm{SO}}_{q,t}(\lambda)/\sqrt{g_\lambda}$
in~\cite{Aganagic:2012au}, where $g_\lambda$ denotes the norm of the
polynomials there. It is now easy to verify that our result coincides with this partition function, up
to the overall factor $\left((q;q)_\infty (t;q)_\infty\right)^{n(g-1)}$.

%=====================================%
\section{Partially Closed Punctures}\label{partclosedpunctures}
%=====================================%
%
In this section we discuss more general theories with partially closed punctures.

%=====================================%
\subsection{Classification}\label{classification}
\begin{figure}
\centering
\includegraphics[height=2.5cm]{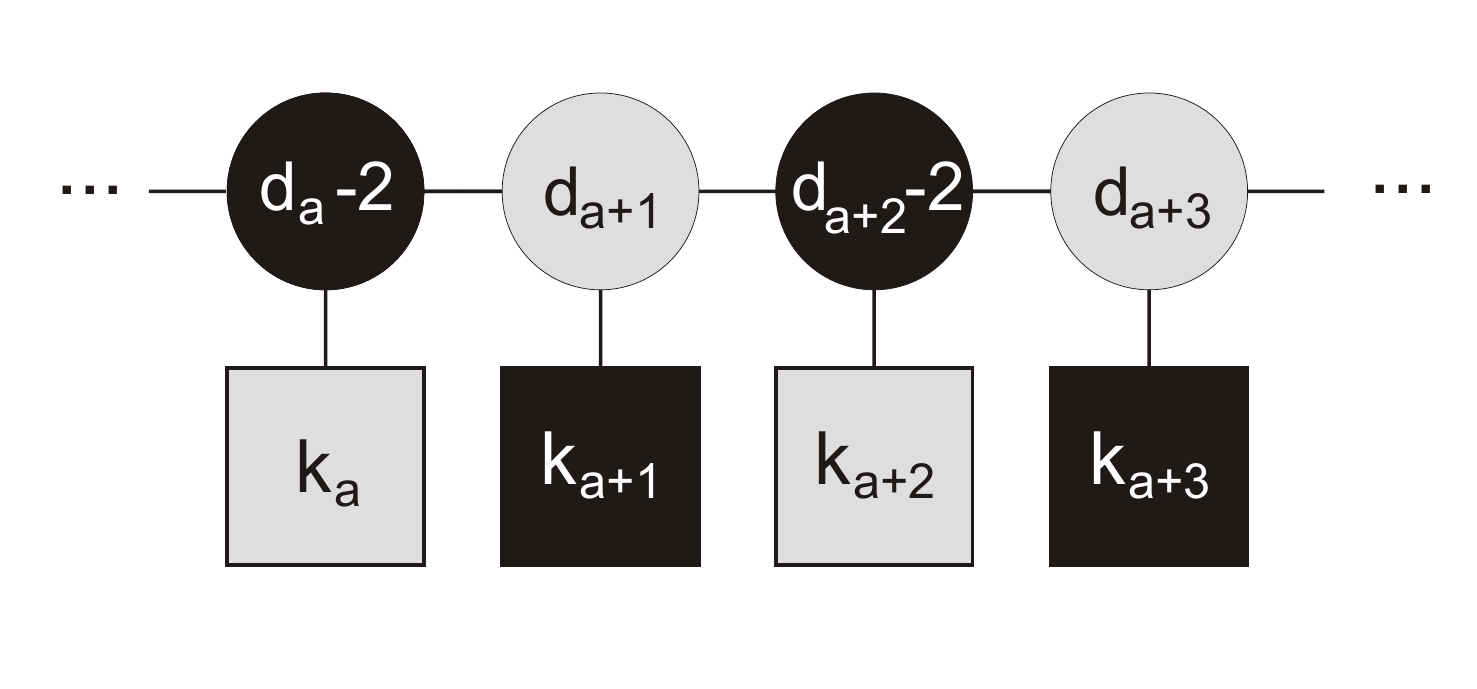}
\caption{Generic alternating quiver.}
\label{tail}
\end{figure}

%=====================================%
The classification of superconformal tails for type $D$ theories was discussed in
\cite{Tachikawa:2009rb}. Let us briefly summarize the main points.
One naturally considers an
alternating quiver of $\mathrm{SO}$ and $\mathrm{USp}$ gauge groups
as in figure~\ref{tail}.
In such a quiver, the requirement that every coupling is
marginal implies that
\begin{align}\label{kddd}
k_a = 2d_a - d_{a+1} - d_{a-1} = (d_a -  d_{a-1}) - (d_{a+1} - d_a).
\end{align}
Since $k_a$ is a non-negative integer one finds that a general
quiver necessarily has
\begin{align}
d_1 < d_2 < \ldots < d_l = \ldots = d_r > d_{r+1} > \ldots > d_m\,.
\end{align}
The part to the left of gauge group $l$ and to the right of gauge
group $r$ are the two tails of this quiver. We denote $d_l = \ldots
= d_r = 2n$ and introduce $d_0 = d_{m+1} = 0.$ Focusing on the right
tail $2n = d_r> d_{r+1}>\ldots > d_m$ and observing that from
(\ref{kddd}) $d_a-d_{a+1}$ is monotonically non-decreasing, we can
associate a Young tableau to this quiver as follows. If the quiver
ends in a $\mathrm{USp}$ gauge group, one considers a grey - {\it i.e.}
$\mathrm{SO}$ -  tableau with length of rows given by $d_r -
d_{r+1},\ d_{r+1}-d_{r+2},\ldots,$ which has a total number of boxes
of $2n.$ If on the other hand the quiver ends in a $\mathrm{SO}$
gauge group, then one can construct a black tableau with length of
rows given by $d_r - d_{r+1},\ d_{r+1}-d_{r+2},\ldots,
d_{m-1}-d_{m},\ d_{m}-d_{m+1}-2,$ which has a total number of boxes equal
to $2n-2.$

The Young tableau thus constructed encodes as usual the embedding of $\mathrm{SU}(2)$ in $\mathrm{SO}(2n)$ $(\mathrm{USp}(2n-2)),$ the commutant of which captures the flavor symmetry information. The embedding is given by the decomposition of the vector representation $2n$ of $\mathrm{SO}(2n)$ (fundamental representation $2n-2$ of $\mathrm{USp}(2n-2)$) under $\mathrm{SU}(2).$ Let us denote the number of columns of height $h$ by $l_h.$ Then the relation between the Young tableau and the decomposition is given for a grey tableau by
\begin{align}\label{decomp2N}
2n \rightarrow \underbrace{1+1+\ldots +1}_{l_1} + \underbrace{2+2+\ldots +2}_{l_2} + \ldots .
\end{align}
Note that the reality of the vector representation requires that $l_h$ is even for even $h.$ The flavor symmetry associated to such a tail is then
\begin{align}
\prod_{h:odd, l_h\geq2} \mathrm{SO}(l_h) \times \prod_{h:even, l_h\geq2} \mathrm{USp}(l_h),
\end{align}
where the $\mathrm{USp}$ groups have an even argument indeed. Following \cite{2011arXiv1106.5410C} we call decompositions (\ref{decomp2N}) in which only even dimensional representations appear ``very even''. Since even dimensional representations must occur with even multiplicities, this case only occurs for even $n.$ Actually, such tableaux correspond to two different punctures which are exchanged by the $\mathbb{Z}_2$ outer automorphism and were colored red and blue in \cite{2011arXiv1106.5410C}.

Note that if the decomposition of the vector representation into $\mathrm{SU}(2)$ representations is of the form $2n \rightarrow (2n - k) +k$ for arbitrary odd $k,$ or $k=n$ if $n$ even, the above algorithm will give rise to a ``$\mathrm{USp}(0)$'' gauge group. The occurrence of such a gauge group can be understood from the brane picture, see \cite{Tachikawa:2009rb}.

For a black tableau one has
\begin{align}
2n-2 \rightarrow \underbrace{1+1+\ldots +1}_{l_1} + \underbrace{2+2+\ldots +2}_{l_2} + \ldots .
\end{align}
Note that the pseudo-reality of the fundamental representation requires that $l_h$ is odd for even $h.$ The flavor symmetry associated to such a tail is then
\begin{align}
\prod_{h:odd, l_h\geq2} \mathrm{USp}(l_h) \times \prod_{h:even, l_h\geq2} \mathrm{SO}(l_h) .
\end{align}

The curve corresponding to these superconformal tails has
$m-r+4$ punctures (or simply said, the number of gauge groups plus
three) of which one is a maximal $\mathrm{SO}$ puncture, one
puncture is the partially closed one under consideration and the
other $m-r+2$ punctures are empty $\mathrm{USp}$  punctures
$\mathbf{\times}.$ Note that a possible ``$\mathrm{USp}(0)$'' gauge
group is also included in this counting of gauge groups.

%
%=====================================%
\subsection{The index  with partially closed punctures}
%=====================================%
%
Following \cite{Gadde:2011uv}, we generalize the structure of the superconformal index in section \ref{Dntheories} to include partially closed punctures. For a three punctured sphere with punctures labeled by a Young tableau $Y_I$ we propose schematically
\begin{align}
\mathcal{I}=\mathcal{A}(q,t)\sum_{\lambda} \f{\prod_{I=1}^{3}
\mathcal{K}(Y_I)P^{\lambda}_{M}\left(Y_I\mid q,t\right)}
{P^{\lambda}_{M\,\mathrm{SO}}\left(1,t,\ldots,t^{n-1} \mid q,t\right)}.
\end{align}
Here the color of the Young tableau determines whether one considers
$\mathrm{SO}$ or $\mathrm{USp}$ polynomials and corresponding
$\mathcal{K}$-factors. Due to the twist line there are either no or
two $\mathrm{USp}$ punctures. If $\mathrm{USp}$ punctures are
present the sum is over $\mathrm{USp}$ representations and the
occurring $\mathrm{SO}$ representations are restricted in the sense
discussed above. The denominator is
always the $(q,t)$- dimension of type $\mathrm{SO}.$

Given this structure, we need to determine for each allowed Young tableau which fugacities one needs to plug in the polynomials and what the $\mathcal{K}$-factor associated to that puncture is.
%
%=====================================%
\subsubsection{Fugacities corresponding to Young diagram}\label{fugacitiesY}
\begin{figure}
\centering
\includegraphics[width=2.6cm]{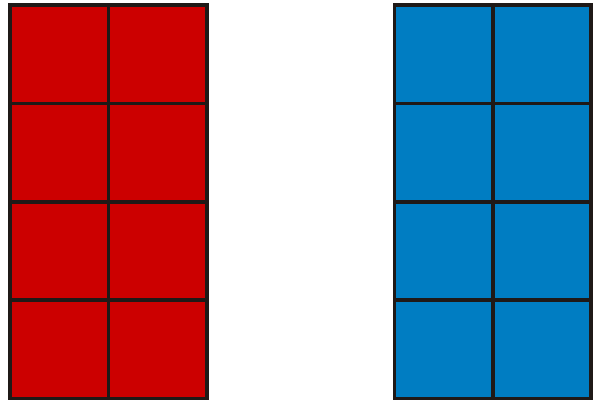}
\caption{An example of a ``very even'' $\mathrm{SO}(8)$ puncture with $\mathrm{USp}(2)$ flavor symmetry. We write $\sum_{i=1}^{4} a_i+a^{-1}_i=(\tau^3+\tau+\tau^{-1}+\tau^{-3})(b+1/b)$, where $b$ is the $\mathrm{USp}(2)$ fugacity. The two essentially inequivalent choices are $a_1= b \tau^3$, $a_2=b \tau$, $a_3=b/\tau$, $a_4=b/\tau^3$, and $a_1= b \tau^3$, $a_2=b \tau$, $a_3=b/\tau$, $a_4=\tau^3/b$. They correspond to the red and blue punctures respectively.~\protect\footnotemark}
\label{redblue}
\end{figure}

%=====================================%
%
To find which fugacities appropriately describe a partially closed
puncture we use that a partially closed $\mathrm{SO}(2n)$
($\mathrm{USp}(2n-2)$) puncture is classified by an embedding of
$\mathrm{SU}(2)$ determined by the decomposition of the vector
(fundamental) representation of this group and that the
multiplicities with which the different $\mathrm{SU}(2)$
representations occur encode the corresponding flavor symmetry. The
fugacities are then determined by looking at the
following equality. On one side of the equality, one simply writes
the character of the vector (fundamental) representation in terms of
its $n$ ($n-1$) fugacities. On the other side of the equality we
write its decomposition determined by the puncture in $\mathrm{SU}(2)$
characters with as $\mathrm{SU}(2)$ fugacity $\tau = t^{1/2},$ where we
replace the multiplicities by the character of the
vector/fundamental of the flavor symmetry determined by that
multiplicity. One can now simply read off the relevant fugacities.
How one identifies the fugacities does not matter in $\mathrm{USp}$, {\it i.e.} neither the
order nor if one chooses $X$ or $X^{-1}$ is relevant, because these operations precisely correspond to Weyl symmetries, under which the Macdonald polynomials are invariant. For the $\mathrm{SO}$ punctures, the order of the fugacities again corresponds to a Weyl symmetry. However, in this case Weyl invariance only means we can invert fugacities in pairs. For odd rank, the tableau always contains at least two columns of odd height. This implies that the two apparently different choices one obtains after exploiting the Weyl symmetries correspond to a renaming of the fugacities. For even rank, there also are odd height columns if the decomposition is not ``very even''. Then again, there is a unique choice of fugacities up to a renaming. However, for the ``very even'' case, there are two essentially inequivalent choices. These are interchanged by the $\mathbb{Z}_2$ outer-automorphism, and they correspond to the red and blue punctures. We illustrate this in figure~\ref{redblue}.
\footnotetext{We checked this assignment by computing the ratio of the index of a free half-hypermultiplet in representation $(1,4,8_s) +(2,1,8_c)$ of the flavor symmetry group $\mathrm{USp}(2)\times \mathrm{USp}(4)\times \mathrm{SO}(8)$ and the index of a free half-hypermultiplet in representation $(1,4,8_c) +(2,1,8_s)$ \cite{2011arXiv1106.5410C}, in a $\tau$ expansion in the Hall-Littlewood limit.}
This distinction disappears once one considers the puncture as part of a Riemann surface containing also $\mathrm{USp}$ punctures, since then the sum over $\mathrm{SO}(2n)$ representations is restricted due to ``diagonality'', {\it i.e.} only representations for which the last orthogonal weight is zero ($\ell_n = 0$) appear. In this case inverting any fugacity becomes a symmetry of the polynomial.
\begin{figure}
\centering
\includegraphics[width=5cm]{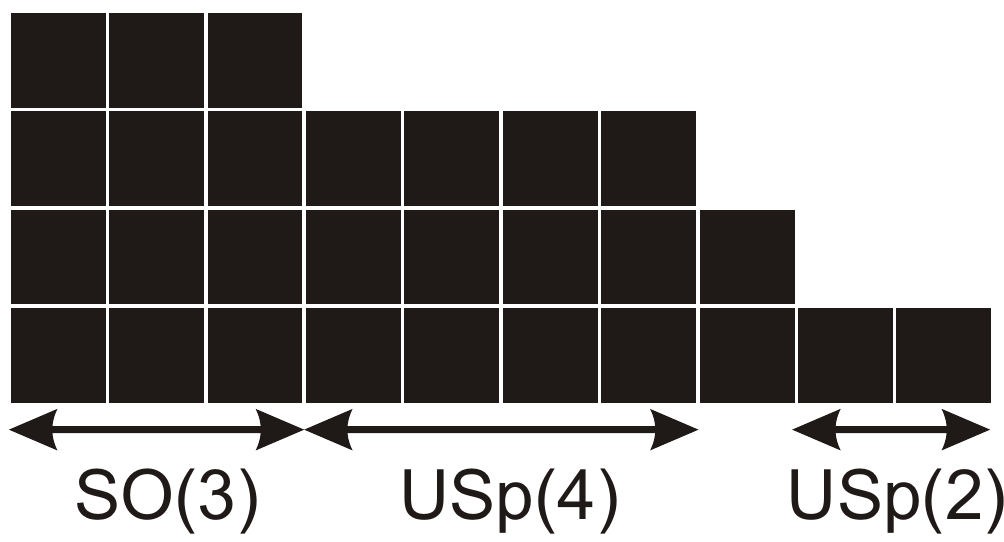}
\caption{$\mathrm{USp}(28)$ puncture with $\mathrm{SO}(3)\times\mathrm{USp}(4)\times\mathrm{USp}(2)$ flavor symmetry.}
\label{USppuncture}
\end{figure}

For example, consider the $\mathrm{USp}(28)$ puncture depicted in figure~\ref{USppuncture}, the flavor symmetry is $\mathrm{SO}(3)\times\mathrm{USp}(4)\times\mathrm{USp}(2)$. Then we write
\begin{align}
\sum_{\alpha=1}^{14} \left( b_\alpha + \frac{1}{b_\alpha} \right) = &\left(\tau^3 + \tau +\tau^{-1} + \tau^{-3} \right) (a + \frac{1}{a} + 1)\notag \\
&+ \left(\tau^2 + 1 +\tau^{-2} \right) \left(\beta_1 + \frac{1}{\beta_1} + \beta_2 + \frac{1}{\beta_2} \right)\notag \\
& + \left( \tau +\tau^{-1} \right)\cdot 1 + 1 \cdot \left(\gamma + \frac{1}{\gamma}\right),
\end{align}
where $a$ is an $\mathrm{SO}(3)$ fugacity, $\beta_1, \beta_2$
are $\mathrm{USp}(4)$ fugacities and $\gamma$ is a $\mathrm{USp}(2)$
fugacity and one takes for example $b_1 = a\tau^3, b_2 = a\tau, b_3
= a\tau^{-1}, b_4 = a\tau^{-3}, b_5 = \tau^3, b_6 = \tau, b_7 =
\beta_1 \tau^2, b_8 = \beta_1, b_9 = \beta_1 \tau^{-2}, b_{10} =
\beta_2 \tau^2, b_{11} = \beta_2, b_{12} = \beta_{2} \tau^{-2},
b_{13}=\tau, b_{14}=\gamma.$
%
%=====================================%
\subsubsection{$\mathcal{K}$-factors}\label{kfact}
%=====================================%
%
We now turn attention to the $\mathcal{K}$-factors, which are independent of the red and blue coloring. For a general
maximal puncture, the $\mathcal{K}$-factor is given in
(\ref{K-factor}) in terms of the roots. For the empty $\mathrm{USp}$
puncture $\mathbf{\times}$, the $\mathcal{K}$-factor was given in
general in (\ref{KfactorminimalUSP}). For the L-shaped empty
$\mathrm{SO}$ puncture, one can easily find the $\mathcal{K}$-factor
by demanding that one obtains a delta function with respect to the propagator measure by closing one puncture
in $T_{SO}$. The resulting $\mathcal{K}$-factor is then
\begin{align}
\mathcal{K} \left(\vcenter{\hbox{\includegraphics[width=0.5cm]{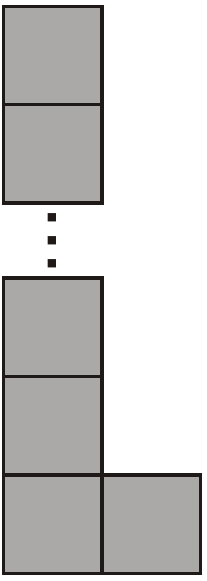}}}\right) = \frac{1}{\mathcal{A}(q,t)}
\end{align}
Note that in general, there are more punctures in both $\mathrm{SO}$
and $\mathrm{USp}$ which do not carry flavor symmetry. All of these factors can be
determined by the following procedure.

As was
discussed above, each puncture is constructed as a certain
superconformal tail and as such is naturally associated to a certain
Riemann surface. By comparing the
index of the tail, which is straightforward to write down since it
is Lagrangian, with the index of the Riemann surface associated to it,
one can fix, in principle, all $\mathcal{K}(Y_I)$.

The index of a Riemann surface of genus $g$ and with
$s$ punctures, labeled by Young tableaux $Y_I$ is given by
\beq
\mathcal{I}_{g,s}=\sum_{\lambda} \left( \f{\mathcal{A}(q,t)}{\mathrm{dim}^{\mathrm{SO}}_{q,t}(\lambda)} \right)^{2g-2+s} \prod_{I=1}^s \mathcal{K}(Y_I)P^{\lambda}_{M}\left(Y_I \mid q,t\right)\,.
\eeq
As before, the color of the $Y_I$ encodes whether the Macdonald polynomials correspond to $\mathrm{SO}$ or $\mathrm{USp}$ groups. Here the sum over $\lambda$ is
restricted as soon as $\mathrm{USp}$ punctures appear, or there is a twist line wrapping a cycle.
Riemann surfaces associated to partially closed punctures have genus zero, one maximal $\mathrm{SO}$ puncture, several empty $\mathrm{USp}$ punctures and the partially closed puncture itself.

To implement the above construction, 
 it is convenient to consider a general Ansatz for the
$\mathcal{K}$-factors, as follows.
 The $\mathcal{K}$-factor for the maximal puncture
diverges if one plugs in the fugacities associated to a non-maximal
Young tableau. However, from~\cite{Gaiotto:2012xa, Gaiotto:2012uq}
we expect to obtain the $\mathcal{K}$-factor associated with a Young
tableau from $\mathcal{K}$ for the maximal puncture by removing some
factors, including the divergent ones, and multiplying by an
appropriate power of $\mathcal{I}^V=(t;q)_\infty (q;q)_\infty$.
Taking out the divergent factors, and allowing for an extra
arbitrary power of $(t;q)_\infty (q;q)_\infty$ thus provides an
Ansatz for each $\mathcal{K}(Y_I)$, and one can determine which
terms must be removed by comparing the index of the superconformal
tail described above to the corresponding Riemann surface.

Equipped with a general recipe,
it would be interesting to check the dualities of \cite{2008JHEP...01..074A}
by comparing the index in different frames, especially for the cases where
enhanced flavor symmetries appear.

We now discuss some simple examples.

%
%=====================================%
\paragraph{$D_3$ theory.}
%=====================================%
%
For the $D_3$ theory all the $\mathcal{K}$-factors for $\mathrm{SO}$
punctures are known from the corresponding $\mathrm{SU}(4)$
punctures, if one appropriately identifies the fugacities
corresponding to the reduced flavor symmetries. Using the above
procedure, these results were also checked in a $q$ ($t$) expansion
for the Schur (Hall-Littlewood) limit,  and thus serve as a proof of
principle. Here $a$, $b$, $c$ denote respectively $\mathrm{SO}(2)$,
$\mathrm{USp}(2)$ and $\mathrm{SO}(3)$ fugacities.
\bea
\mathcal{K}\left(\vcenter{\hbox{\includegraphics[width=0.9cm]{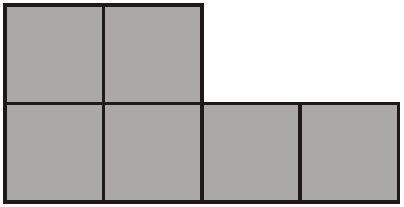}}}\right)&=&
\f{1}{(q;q)_\infty^{3/2}(t;q)_\infty^{1/2}(t^2,t^{3/2}(b/a)^{\pm},t^{3/2} (b a)^{\pm},t b^{\pm 2};q)_\infty}\,,\nn\\
\mathcal{K}\left(\vcenter{\hbox{\includegraphics[width=0.9cm]{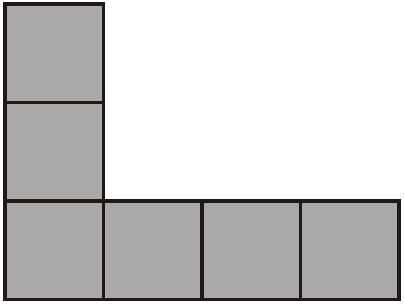}}}\right)&=&
\f{(t;q)_\infty^{1/2}}{(q;q)_\infty^{3/2}(t^2,q)_\infty^2(t c^{\pm },t^2 c^{\pm };q)_\infty}\,,\nn\\
\mathcal{K}\left(\vcenter{\hbox{\includegraphics[width=0.5cm]{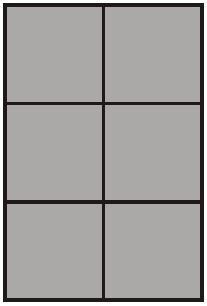}}}\right)&=&
\f{(t;q)_\infty^{1/2}}{(q;q)_\infty^{3/2}(t^2,t^3,t^2 a^{\pm 2 };q)_\infty}\,.
\eea
The three-punctured sphere consisting of the second
puncture (which has $\mathrm{SO}(3)$ flavor symmetry) and two
maximal punctures gives the $\mathrm{E}_7$ SCFT of
Minahan-Nemeschansky \cite{Minahan:1996cj}, and its index can be
independently (and in agreement with the above expression) computed
by using the S-duality described in \cite{Tachikawa:2009rb}.

\begin{figure}
\centering
\subfloat[]{\label{SO:SO2} \includegraphics[width=3.5cm]{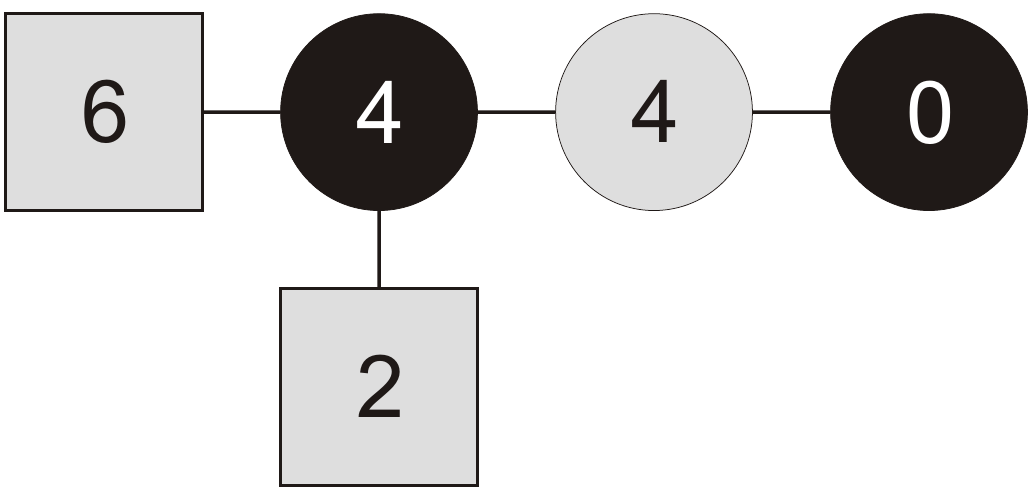}}   \quad \quad 
\subfloat[]{\label{SO:SO2G} \includegraphics[width=3cm]{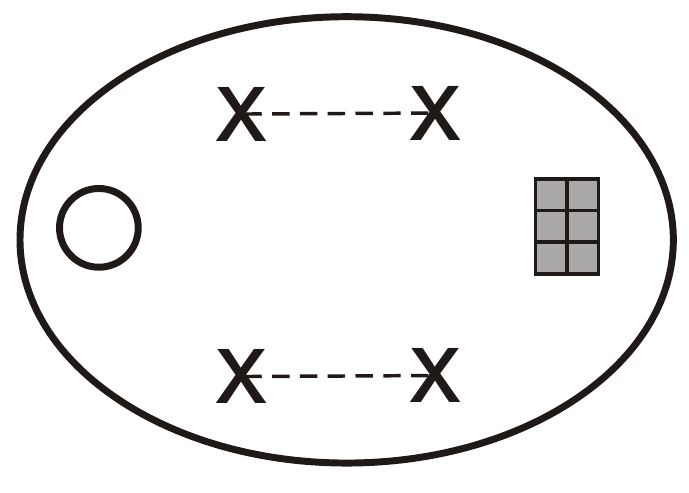}}  \quad \quad \quad
\subfloat[]{\label{USp:SO2}\includegraphics[width=2.5cm]{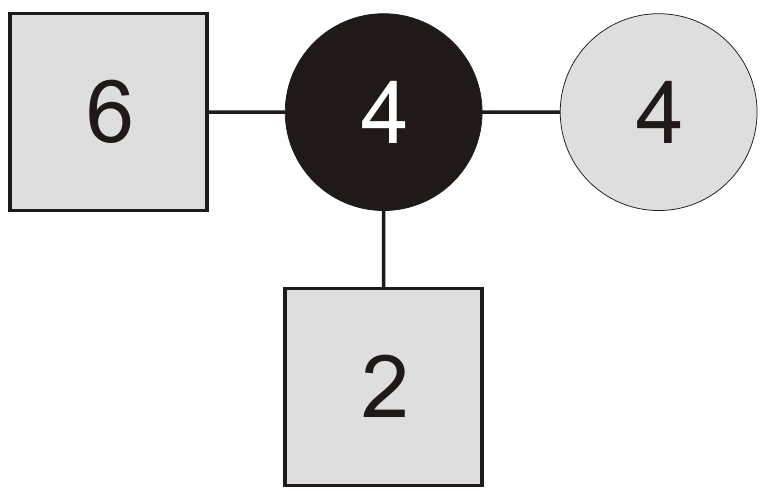}}   \quad \quad
\subfloat[]{\label{USp:SO2G} \includegraphics[width=3cm]{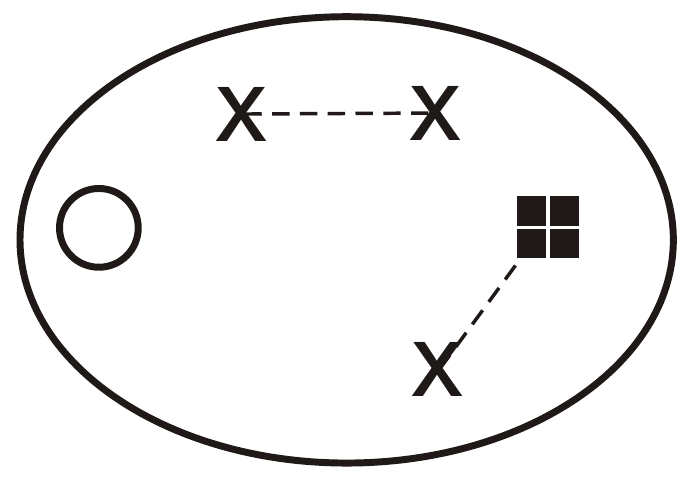}} 
\caption{Superconformal tail and corresponding Riemann surface for the $D_3$ $\mathrm{SO}$ puncture with  $\mathrm{SO}(2)$ flavor symmetry a) and b), and for the $D_3$ $\mathrm{USp}$ puncture with  $\mathrm{SO}(2)$ flavor symmetry c) and d).}
\end{figure}

Note that the tail corresponding to the last puncture
($\mathrm{SO}(2)$ flavor symmetry) ends on a ``$\mathrm{USp}(0)$''
gauge group, as shown in figure~\ref{SO:SO2}, which also shows the
corresponding Riemann surface \ref{SO:SO2G}. This happens for the class of punctures
described in section~\ref{classification}, but the procedure
outlined above is still applicable.
We find $\mathcal{K}$ for the $\mathrm{USp}$ punctures with
$\mathrm{USp}(2)$ and $\mathrm{SO}(2)$ flavor symmetries to be
respectively given by
\bea
\mathcal{K}\left(\vcenter{\hbox{\includegraphics[width=0.75cm]{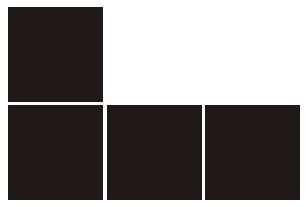}}}\right)&=&
\f{1}{(q,t^2,t b^{\pm 2},t^{3/2} b^{\pm};q)_\infty}\,,\nn\\
\mathcal{K}\left(\vcenter{\hbox{\includegraphics[width=0.5cm]{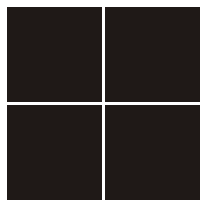}}}\right)&=&
\f{1}{(q,t^2,t^2 a^{\pm 2};q)_\infty}\,.
\eea
Here $b$ denotes an  $\mathrm{USp}(2)$ and $a$ an $\mathrm{SO}(2)$
fugacity. These factors were checked in the Schur and
Hall-Littlewood limits by comparing the index of the tail with the
one for the curve, in an expansion in $q$ and $\tau$ respectively.
The superconformal tail for the $\mathrm{SO}(2)$ flavor symmetry is
shown in figure~\ref{USp:SO2}, along with the corresponding curve \ref{USp:SO2G}. The index for this linear quiver is the same as the index for the one
shown in figure~\ref{SO:SO2}, since the ``$\mathrm{USp}(0)$'' gauge
group does not contribute to the index. From the curve this
implies that (after integrating against the $\mathrm{SO}(6)$ polynomials)
\small
\bea
&&\f{\mathcal{A}(q,t)}{\mathrm{dim}^{\mathrm{SO}}_{q,t}((\lambda_1,\lambda_2,\lambda_2))}
 \mathcal{K}\left(\vcenter{\hbox{\includegraphics[width=0.5cm]{SO2.pdf}}}\right) P^{(\lambda_1,\lambda_2,\lambda_2)}_{M\,\mathrm{SO}}\left(a t,\f{a}{t},a\mid q,t\right)
\mathcal{K}_{\mathrm{USp}}(\mathbf{\times})P^{(\lambda_1,\lambda_2)}_{M\,\mathrm{USp}}\left(t^{1/2},t^{3/2}\mid q,t\right)=\nn\\
&&=
\mathcal{K}\left(\vcenter{\hbox{\includegraphics[width=0.5cm]{SO2-_USp_.pdf}}}\right)
P^{(\lambda_1,\lambda_2)}_{M\,\mathrm{USp}}\left(a \sqrt{t},\f{a}{\sqrt{t}}\mid q,t\right)\,,
\eea
\normalsize
where $\lambda_1$ and $\lambda_2$ are Dynkin labels. This identity is true for any Dynkin labels $\lambda_1$ and $\lambda_2$, and so taking $\lambda_1=\lambda_2=0$ the representation dependent part cancels on its own and we get a relation between the $\mathcal{K}$-factors,
\beq
\mathcal{A}(q,t)
 \mathcal{K}\left(\vcenter{\hbox{\includegraphics[width=0.5cm]{SO2.pdf}}}\right) 
\mathcal{K}_{\mathrm{USp}}(\mathbf{\times})=
\mathcal{K}\left(\vcenter{\hbox{\includegraphics[width=0.5cm]{SO2-_USp_.pdf}}}\right)\,.
\eeq
We also get a relation between the $\mathrm{SO}$ and $\mathrm{USp}$ polynomials. In general, any $\mathrm{SO}$ puncture labeled by a Young diagram with only two columns can similarly be related to a $\mathrm{USp}$ puncture labeled by the Young diagram one obtains by removing the bottom two boxes of the $\mathrm{SO}$ Young diagram and coloring it black, thus giving a relation between the $\mathcal{K}$-factors and an identity involving Macdonald polynomials.

\paragraph{$D_2$ theory.} Another simple example of this relation, in $D_2,$ is given by the colored box-shaped $\mathrm{SO}$ puncture with $\mathrm{USp}(2)$ flavor symmetry. The $\mathcal{K}$-factor for this puncture is
\begin{align}
\mathcal{K}\left(\vcenter{\hbox{\includegraphics[width=0.5cm]{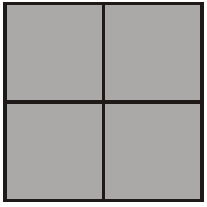}}}\right) = \frac{1}{\left(q,t^2,ta^{\pm 2} ; q \right)_\infty},
\end{align}
which satisfies 
\beq
\mathcal{A}(q,t)
 \mathcal{K}\left(\vcenter{\hbox{\includegraphics[width=0.5cm]{boxUSp2.pdf}}}\right) 
\mathcal{K}_{\mathrm{USp}}(\mathbf{\times})=
\mathcal{K}\left(\vcenter{\hbox{\includegraphics[width=0.5cm]{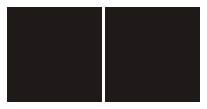}}}\right)\,.
\eeq
We also have
\beq
\f{1}{\mathrm{dim}^{\mathrm{SO}}_{q,t}((\lambda,\lambda))}
P^{(\lambda,\lambda)}_{M\,\mathrm{SO}}\left(a t^{1/2},\f{a}{t^{1/2}}\mid q,t\right)
P^{(\lambda)}_{M\,\mathrm{USp}}\left(t^{1/2}\mid q,t\right)=
P^{(\lambda)}_{M\,\mathrm{USp}}\left(a\mid q,t\right)\,.\label{boxD2}
\eeq
At this point we can return to the $A_1$ interpretation of $D_2$ theories and include this partially closed puncture. Namely, this partial closing of the $\mathrm{SO}(4)$ puncture amounts to fully closing one of the two $\mathrm{SU}(2)$ punctures in which it decomposes in $A_1$ language. The coloring is simply encoded in the choice of which $\mathrm{SU}(2)$ becomes fully closed.
More in detail, one has for the red puncture
\begin{align}\label{PMred}
P_{M\, \mathrm{SO}}^{(\lambda_1,\lambda_2)}\left(a t^{1/2},\f{a}{t^{1/2}}\mid q,t\right) = P_{M\, \mathrm{SU}}^{(\lambda_1)}\left(t^{1/2},t^{-1/2} \mid q,t\right)P_{M\, \mathrm{SU}}^{(\lambda_2)}\left(a,1/a \mid q,t\right),
\end{align}
and for the blue puncture 
\begin{align}\label{PMblue}
P_{M\, \mathrm{SO}}^{(\lambda_1,\lambda_2)}\left(a t^{1/2},\f{t^{1/2}}{a}\mid q,t\right) = P_{M\, \mathrm{SU}}^{(\lambda_1)}\left(a,1/a \mid q,t\right)P_{M\, \mathrm{SU}}^{(\lambda_2)}\left(t^{1/2},t^{-1/2} \mid q,t\right).
\end{align}
Note that when one imposes ``diagonality'', the coloring becomes as expected irrelevant. For the (color-independent) $\mathcal{K}$-factor one has
\begin{align}
\mathcal{K}\left(\vcenter{\hbox{\includegraphics[width=0.5cm]{boxUSp2.pdf}}}\right)  = \mathcal{K}_{\mathrm{SU}}\left(\vcenter{\hbox{\includegraphics[width=0.5cm]{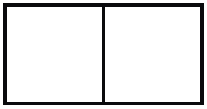}}}\right) \mathcal{K}_{\mathrm{SU}}\left(\vcenter{\hbox{\includegraphics[width=0.25cm]{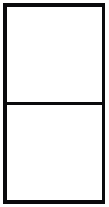}}}\right)\,.
\end{align}
Also note that when using \eqref{PMred} (or \eqref{PMblue}), rewriting the Macdonald dimension in terms of $\mathrm{SU}$ Macdonald dimensions using \eqref{factorize}, and naturally thinking of the $\mathrm{USp}(2)$ polynomials in terms of $\mathrm{SU}(2)$ polynomials, equation \eqref{boxD2} is trivial.

\acknowledgments 
We are grateful to
Abhijit Gadde, Shlomo Razamat and Wenbin Yan for useful discussions and comments.
This work is partially supported by the NSF under Grants PHY-0969919 and PHY-0969739. Any opinions, findings, and conclusions
or recommendations expressed in this material are those of the authors and do not necessarily
reflect the views of the National Science Foundation. The research of
ML is partially funded by FCT - Portugal through grant SFRH/BD/70614/2010.

%
%=====================================%
\appendix
%=====================================%

%=====================================%
\section{Macdonald Polynomials and Macdonald Operator}\label{sectionMacdonaldpolyandoperator}
%=====================================%
%
In this appendix we review the results from~\cite{Macdonald} that are relevant for our purposes.
%
%=====================================%
%\subsection{Some notation}
\subsection{Weyl invariant polynomials and the Macdonald operator}
%=====================================%
Let us start by summarizing some notational conventions, for which
we follow~\cite{Macdonald}. Consider a Lie algebra $\mathfrak{g}.$
We denote the root system of $\mathfrak{g}$ by $R$ and the system of
positive roots by $R^+.$ For every root $\alpha$, we denote the
coroot as $\alpha^{\vee}\equiv 2\alpha / \left\langle \alpha,
\alpha\right\rangle.$ The root lattice, spanned by the simple roots
$\{\alpha_i\ |\  1\leq i \leq \text{rank } \mathfrak{g} = n\}$ is
denoted by $Q,$ and its positive part by $Q^+.$ Finally, the weight
lattice is denoted by $P = \left\{\lambda \in \mathbb{R}^n\ |\
\left\langle \lambda, \alpha^{\vee}\right\rangle \in \mathbb{Z}
\right\},$ and by $P^+$ we denote the dominant weights, {\it i.e.}
$P^+=\left\{ \lambda \in P \ |\  \forall \alpha \in R^+:
\left\langle \lambda, \alpha^{\vee}\right\rangle \in \mathbb{N}
\right\}$. A basis for $P^+$ is given by the fundamental weights
$\omega_i.$ The components of an arbitrary weight $\lambda$ in this
basis will be denoted $\lambda_i,$ {\it i.e.} $\lambda = \sum_i \lambda_i
\omega_i.$ The $\lambda_i$ are called the Dynkin lables of the
representation. Another basis is given by the standard orthonormal
basis, denoted by $\epsilon_i.$ The components in
this basis will be denoted $\ell_i.$ On the weight lattice we can
introduce a partial order as $\lambda \geq \mu \Leftrightarrow
\lambda - \mu \in Q^+.$

The Weyl group $W$ is a finite group generated by the simple Weyl reflections $\sigma_\alpha,$ for all roots $\alpha.$ These are defined on any weight in $P$ as
\begin{align}
\sigma_\alpha(\lambda) = \lambda - \left\langle \lambda, \alpha^{\vee}\right\rangle \alpha.
\end{align}

The group algebra of the weight lattice $P$ is denoted by $A.$ It is
generated by the formal exponentials $e^\lambda,$ satisfying
$e^\lambda e^\mu = e^{\lambda + \mu}$ and
$\left(e^\lambda\right)^{-1} = e^{-\lambda}.$  We can formally
identify the variables $x_i$ as $x_i = e^{\epsilon_i}.$ The action
of the Weyl group on $P$ is uplifted to $A$ as $w(e^\lambda) =
e^{w\lambda},$ where $w\in W.$ The subalgebra of $A$ invariant under
$W$ is denoted by $A^W,$ and is most easily spanned by the symmetric
orbit-sums $\{m_{\lambda}\ |\ \lambda \in P^+ \}$. These are defined
as $m_{\lambda} = \sum_{\mu \in W(\lambda)} e^\mu,$ where
$W(\lambda)$ denotes the Weyl-orbit of $\lambda.$ Another well-known
basis for $A^W$ is given by the group characters $\chi_{\lambda}$
which are given by the Weyl character formula:
\begin{align}
\chi_{\lambda} = \frac{\sum_{w\in W} \epsilon(w)e^{w(\lambda +
\rho)}}{\sum_{w\in W} \epsilon(w)e^{w\rho}},
\end{align}
where $\rho = \frac{1}{2} \sum_{\alpha \in R^+}\alpha$ is the Weyl
vector and $\epsilon(w)$ is the signature of $w\in W.$ The signature
is defined as $\epsilon(w)=(-)^{l(w)}$ where $l(w)$ is the minimum
number of simple Weyl reflections in which $w$ can be decomposed.

An alternative characterization of the Weyl characters is as the
unique polynomials satisfying two conditions, namely that they can
be written in terms of the orbit-sums for some coefficients
$K_{\lambda \mu}$ as
\begin{align}\label{Weylcharacter}
\chi_{\lambda} = m_{\lambda} + \sum_{\mu \in P^+,\ \mu<\lambda}
K_{\lambda \mu} m_{\mu},
\end{align}
which constrains the leading behavior, and that they are orthogonal
under the Haar measure\footnote{See below for the general discussion
of the inner product.}. It is these two criteria which are
generalized to define the Hall-Littlewood and Macdonald polynomials
for arbitrary root systems.

Let us start by introducing a 2-parameter generalization of the Haar
measure
\begin{align}
\Delta_{M}(q,t) = \prod_{\alpha \in R}
\frac{(e^\alpha;q)_{\infty}}{(te^\alpha;q)_{\infty}},
\label{MacMeasure}
\end{align}
where we used the q-Pochhammer symbols $(a;q)_{\infty} =
\prod_{j=0}^{\infty}(1-aq^j).$ We will call this measure the
Macdonald measure. For $t=q$ it reduces to the Haar measure,
\begin{align}\label{HaarMeasure}
\Delta= \prod_{\alpha \in R} (1-e^\alpha),
\end{align}
and for $q\to 0$ it results in the Hall-Littlewood measure
\begin{align}\label{HLMeasure}
\Delta_{HL}(t) = \prod_{\alpha \in R}
\frac{1-e^\alpha}{1-te^\alpha}.
\end{align}

The definition of the scalar product of two functions $f =
\sum_{\lambda \in P} f_\lambda e^\lambda$ and $g=\sum_{\lambda \in
P} g_\lambda e^\lambda$ is most easily written down if we first
write them in terms of the $x$-variables introduced above. To do so
we write $\lambda = \sum_i \ell_i \epsilon_i$ in the orthogonal
basis, and thus $e^\lambda = \prod_i x_i^{\ell_i}.$ We also
introduce $\bar{g} = \sum_{\lambda \in P} g_\lambda e^{-\lambda} =
\sum_{\lambda \in P} g_\lambda \prod_i x_i^{-\ell_i},$ and the
shorthand notation $\oint[dx] \equiv \oint\left(\prod_i
\frac{dx_i}{2\pi \mathrm{i} x_i} \right).$ Then the inner product is
defined as
\begin{align}\label{innerprod}
\left\langle f, g \right\rangle = \frac{1}{|W|} \oint [dx] f(x)
\bar{g}(x) \tilde{\Delta}(x),
\end{align}
where $f(x)=f(x_1,x_2,\ldots)$ and so forth, $|W|$ is the order of
the Weyl group and $\tilde{\Delta}$ denotes one of the measures
introduced above.

The generalization we were alluding to is now as follows.
There exists a unique basis for $A^W$ of functions $\{\tilde{P}_\lambda\ |\
\lambda \in P^+\}$ such that
\begin{align}\label{Macdonaldpolygeneral}
\tilde{P}_\lambda = m_\lambda + \sum_{\mu \in P^+,\ \mu < \lambda}
u_{\lambda \mu} m_{\mu},
\end{align}
where the coefficients are rational functions of $q$ and $t,$ such
that they are orthogonal under what we called the Macdonald measure.

Note that instead of normalization, one favors a constrained leading
behavior. We reserve the notation $\tilde{P}$ for polynomials satisfying this requirement. In the main text we will use polynomials orthonormal under \eqref{innerprod} with the Macdonald measure \eqref{MacMeasure}, which we denote by $P.$

As mentioned above already, for $q=t$ the polynomials are the Weyl
characters~\eqref{Weylcharacter}. For $q\rightarrow 0,$ one has the
Hall-Littlewood polynomials, which are given explicitly by
\begin{align}\label{HLpolynomials}
\tilde{P}_{\lambda}^{HL} = W_{\lambda}(t)^{-1} \sum_{w\in W}
w\left(e^\lambda \prod_{\alpha \in R^+} \frac{1-t
e^{-\alpha}}{1-e^{-\alpha}}\right),
\end{align}
where
\begin{align}
W_\lambda(t) = \sum_{\substack{w\in W \\  w\lambda = \lambda}}
t^{l(w)}.
\end{align}

For general $t,q,$ a simple expression as above for the
Hall-Littlewood polynomials or Weyl-characters is absent. However,
the polynomials can be generated quite easily through a
determinantal formula~\cite{vanDiejen}. This formula makes use of the proposition that there exists a linear operator $D:A^W \rightarrow A^W$ such that
\begin{enumerate}
  \item $D$ is selfadjoint, {\it i.e.} $\left\langle Df, g \right\rangle =
  \left\langle f, Dg \right\rangle$ for all $f,g \in A^W;$
  \item $D$ is triangular relative to the basis $m_{\lambda},$ {\it i.e.}
  for each $\lambda\in P^+,$ $Dm_{\lambda}$ is of the form
\begin{align*}
  Dm_{\lambda} = \sum_{\mu \leq \lambda} c_{\lambda \mu} m_{\mu};
  \end{align*}
  \item the eigenvalues of $D$ are distinct, {\it i.e.} if $\lambda \neq \mu \in
  P^+$ then $c_{\lambda\lambda} \neq c_{\mu\mu}.$
\end{enumerate}
It is easy to understand how the existence theorem follows from this
proposition. Namely, given an operator $D$ satisfying these three
properties, one can consider for each $\lambda \in P^+$ the
eigenfunction $\tilde{P}_\lambda$ with eigenvalue $c_{\lambda\lambda}.$ One
can normalize this eigenfunction such that the coefficient of
$m_\lambda$ equals 1. Moreover, using the selfadjointness of $D$ and
the nondegeneracy of its eigenvalues, one can argue that
$c_{\lambda\lambda}\left\langle \tilde{P}_\lambda, \tilde{P}_{\mu} \right\rangle =
\left\langle D \tilde{P}_\lambda, \tilde{P}_{\mu} \right\rangle = \left\langle
\tilde{P}_\lambda, D \tilde{P}_{\mu} \right\rangle =c_{\mu\mu}\left\langle
\tilde{P}_\lambda, \tilde{P}_{\mu} \right\rangle,$ which implies that for $\mu \neq
\lambda$ one has $\left\langle \tilde{P}_\lambda, \tilde{P}_{\mu} \right\rangle =
0.$

The proof of the proposition is given in~\cite{Macdonald} and is
simply based on the construction of an operator satisfying the above
three properties. To that purpose one starts by constructing for
each minuscule weight $\pi$ for the dual root system $R^{\vee}$
\footnote{Such a minuscule weight is characterized by the
requirement that $\left\langle \pi,\alpha \right\rangle = 0 \text{
or } 1 $ for all $\alpha \in R^+.$ Note that $E_8,$ $F_4,$ and $G_2$
do not have minuscule weights. They can be dealt with differently.
See~\cite{Macdonald}.} the operator $D_{\pi}$ as
\begin{align}
D_\pi = \frac{1}{|W_{\pi}|}\sum_{w\in W} \left( \prod_{\alpha\in
R^+}\frac{1-t^{\left\langle \pi,\alpha \right\rangle }
e^{w(\alpha)}} {1- e^{w(\alpha)}}\right) T_{w(\pi),q},
\end{align}
where $T_{x,q}$ is defined by its action on the exponentials
$e^{\lambda}$ as
\begin{align}
T_{x,q} e^{\lambda} = q^{\left\langle \lambda,x \right\rangle }
e^{\lambda}.
\end{align}
One can prove that these operators satisfy the requirements 1. and
2. The property 3. is also satisfied except for the $D$-series. In
order to lift the degeneracy of the eigenvalues in the $D$-case, one
constructs an appropriate linear combination of the operators
$D_{\pi_1}$ and $D_{\pi_2},$ where $\pi_1$ and $\pi_2$ are the two
minuscule weights of $R^{\vee}_{D} = R_D,$ namely the fundamental
weights corresponding to the two spinor representations, {\it i.e.}
$\omega_n$ and $\omega_{n-1}$. For any integer $ N
> \frac{1}{2} n(n-1),$ where $n$ is the rank, one considers
\begin{align}\label{MDoperatorDseries}
D_{\mathrm{SO}} = \frac{1}{2}t^{-N} \left( (t^N + 1) D_{\pi_1} +
(t^N - 1 ) D_{\pi_2} \right),
\end{align}
which now also satisfies property 3.

The eigenvalues of the operators $D_{\pi}$ can be written in general
as
\begin{align}
c_{\lambda \lambda}(\pi) = t^{\left\langle \pi, \rho \right\rangle}
\sum_{\tau \in W(\pi)} t^{\left\langle \tau,\rho \right\rangle}
q^{\left\langle \tau, \lambda \right\rangle}.
\end{align}
One thus has for the eigenvalues of $D_{\mathrm{SO}}$ in~
\eqref{MDoperatorDseries}
\begin{align}
c_{\lambda}^{\mathrm{SO}} = \frac{1}{2}t^{-N} \left( (t^N + 1)
t^{\left\langle \pi_1, \rho \right\rangle} \sum_{\tau \in W(\pi_1)}
t^{\left\langle \tau,\rho \right\rangle} q^{\left\langle \tau,
\lambda \right\rangle} + (t^N - 1 ) t^{\left\langle \pi_2, \rho
\right\rangle} \sum_{\tau \in W(\pi_2)} t^{\left\langle \tau,\rho
\right\rangle} q^{\left\langle \tau, \lambda \right\rangle} \right).
\end{align}

%
%=====================================%
\subsection{More explicit expressions for the $C$- and $D$-series}\label{Sec:ExplicitMac}
%=====================================%
%
Let us write the Weyl group, the Macdonald operator and its eigenvalues a little
more explicit.
%
%=====================================%
\subsubsection{The case $C_n = \mathrm{USp}(2n)$}
%=====================================%
%
The Weyl group of $C_n$ is given by all possible permutations and sign changes of the orthogonal weights. The dual root lattice $R^{\vee}$ of $C_n$ equals the root lattice of
$B_n.$ The unique minuscule weight of $R^{\vee}$ is then the
fundamental weight $\omega_n$ of $B_n.$

The Macdonald operator then reads explicitly~\cite{1997q.alg....12054M}
\begin{align}
D_{\mathrm{USp}} \equiv D_{\pi = \omega_n^{(B_n)}} =
\sum_{s_1,\ldots,s_n = \pm 1}\ \prod_{1\leq i<j\leq n} \frac{1-t
x_i^{s_i} x_j^{s_j}}{1- x_i^{s_i} x_j^{s_j}} \prod_{1\leq i \leq
n}\frac{1-t x_i^{2s_i} }{1- x_i^{2s_i} } T_{x_i}^{\frac{s_i}{2}},
\label{MacOpC}
\end{align}
where $T_{x_i}$ is defined as $(T_{x_i}f)(x_1,\ldots,x_n) =
f(x_1,\ldots,q x_i,\ldots,x_n).$ The eigenvalues of the Macdonald operator can be written explicitly
as
\begin{align}
c_{\lambda}^{\mathrm{USp}} \equiv
c_{\lambda\lambda}(\omega_n^{(B_n)}) = \prod_{j =
1}^{n}\left(t^{n+1-j} q^{\ell_j/2} + q^{-\ell_j/2} \right).
\end{align}
%
%=====================================%
\subsubsection{The case $D_n = \mathrm{SO}(2n)$}
%=====================================%
%
The Weyl group is given by all possible permutations and all even number of sign changes of the orthogonal weights. The operator~\eqref{MDoperatorDseries} reads explicitly
\begin{align}
D_{\mathrm{SO}}=\frac{1}{2} \left(  (D_{\pi_1} +D_{\pi_2}) +  t^{-N}
(D_{\pi_1} -D_{\pi_2}) \right), \label{MacOpD}
\end{align}
where
\begin{align}
D_{\pi_1}&= \sum_{\substack{s_1,\ldots,s_{n-1} = \pm 1 \\ s_n =
\prod_{j=1}^{n-1}s_j }}\ \prod_{1\leq i<j\leq n} \frac{1-t x_i^{s_i}
x_j^{s_j}}{1- x_i^{s_i} x_j^{s_j}} \prod_{1\leq i \leq n}
T_{x_i}^{\frac{s_i}{2}},\\
D_{\pi_2}&= \sum_{\substack{s_1,\ldots,s_{n-1} = \pm 1 \\ s_n =
-\prod_{j=1}^{n-1}s_j }}\ \prod_{1\leq i<j\leq n} \frac{1-t
x_i^{s_i} x_j^{s_j}}{1- x_i^{s_i} x_j^{s_j}} \prod_{1\leq i \leq n}
T_{x_i}^{\frac{s_i}{2}}.
\end{align}
Its eigenvalues can be written as
\begin{align}
c_{\lambda}^{\mathrm{SO}} = \frac{1}{2}\prod_{j=1}^{n}\left( t^{n-j}
q^{\ell_j/2} + q^{-\ell_j/2} \right) + \frac{1}{2} t^{-N}
\prod_{j=1}^{n}\left( t^{n-j} q^{\ell_j/2} - q^{-\ell_j/2} \right).
\end{align}
It is important to observe here that the eigenvalues of the
representation $\lambda' = \sum_{i=1}^{n-1} \ell_i \epsilon_i$ of
$C_{n-1}$ equal the eigenvalues of the representation $\lambda =
\sum_{i=1}^{n-1} \ell_i \epsilon_i + 0\ \epsilon_n$ of $D_{n}.$ In
Dynkin labels, the equivalent statement is that the eigenvalue of
the representation $\lambda' = \sum_{i=1}^{n-1} \lambda_i \omega_i $
of $C_{n-1}$ equals that of the representation $\lambda =
\sum_{i=1}^{n-1} \lambda_i \omega_i + \lambda_{n-1} \omega_n$ of
$D_n.$

%=====================================%
\section{Interwining Property of the Free Hyper Index }\label{appendixactionmdoperator}
%=====================================%
%
After some elementary algebraic manipulations and using the
properties of the $q$-Pochhammer symbols, one can simplify the action
of the conjugated Macdonald operator on the free half-hypermultiplet
index for $\mathrm{USp}$ to
\begin{align}\label{DUspacting}
&   \hat D_\mathrm{USp} (\mathbf{b}) \mathcal{I}(\mathbf{a},\mathbf{b}) \equiv \mathcal{K}_\mathrm{USp} (\mathbf{b}) D_\mathrm{USp}(\mathbf{b})
\mathcal{K}^{-1}_\mathrm{USp} (\mathbf{b})
\mathcal{I}(\mathbf{a},\mathbf{b})=
\sum _{s_1,\cdots,s_{n-1}=\pm 1} \prod _{\alpha\leq \beta}^{n-1} \frac{1- b_\alpha^{-s_\alpha}b_\beta^{-s_\beta} t/q}{1- b_\alpha^{s_\alpha}b_\beta^{s_\beta}}\times\nn\\
&\times\prod _{i =1}^n \prod _{\alpha=1}^{n-1} \frac{1}{\left(1-\sqrt{t/q} \left(a_{i }/b_\alpha\right)^{s_\alpha}\right)\left(1-\sqrt{t/q} \left(a_{i } b_\alpha\right)^{-s_\alpha}\right)\left(\sqrt{t\,q}\, a_{i }^\pm b_\alpha^\pm ;q\right)_\infty}\,,
\end{align}
and for $\mathrm{SO}$ to
\begin{align} \label{Dpi1acting}
&\hat D_{\pi_1} (\mathbf{a})  \mathcal{I}(\mathbf{a},\mathbf{b}) \equiv  \mathcal{K}_\mathrm{SO} (\mathbf{a}) D_{\pi_1}(\mathbf{a})
\mathcal{K}^{-1}_\mathrm{SO} (\mathbf{a})
\mathcal{I}(\mathbf{a},\mathbf{b})=
\sum _{\substack{s_1,\cdots, s_{n-1}=\pm 1 \\ s_n=\prod_{i=1}^{n-1}s_i}} \prod _{i <j}^{n} \frac{1- a_i^{-s_i}a_j^{-s_j} t/q}{1- a_i^{s_i}a_j^{s_j}}\times\nn\\
&\times\prod _{i =1}^n \prod _{\alpha=1}^{n-1} \frac{1}{\left(1-\sqrt{t/q} \left(b_\alpha/a_{i }\right)^{s_i}\right)\left(1-\sqrt{t/q} \left(a_{i } b_\alpha\right)^{-s_i}\right) \left(\sqrt{t\,q}\, a_{i}^\pm  b_\alpha^\pm ;q\right)_\infty}\,,
\end{align}
and
\begin{align}\label{Dpi2acting}
&\hat D_{\pi_2} (\mathbf{a})  \mathcal{I}(\mathbf{a},\mathbf{b}) \equiv \mathcal{K}_\mathrm{SO} (\mathbf{a}) D_{\pi_2}
\mathcal{K}^{-1}_\mathrm{SO} (\mathbf{a})
\mathcal{I}(\mathbf{a},\mathbf{b})=
\sum _{\substack{s_1,\cdots, s_{n-1}=\pm 1 \\ s_n=-\prod_{i=1}^{n-1}s_i}} \prod _{i <j}^{n} \frac{1- a_i^{-s_i}a_j^{-s_j} t/q}{1- a_i^{s_i}a_j^{s_j}}\times\nn\\
&\times\prod _{i =1}^n \prod _{\alpha=1}^{n-1} \frac{1}{\left(1-\sqrt{t/q}
\left(b_\alpha/a_{i }\right)^{s_i}\right)\left(1-\sqrt{t/q}
\left(a_{i } b_\alpha\right)^{-s_i}\right) \left(\sqrt{t\,q}\,
a_{i}^\pm  b_\alpha^\pm ;q\right)_\infty}\,.
\end{align}
Note that one can obtain (\ref{Dpi2acting}) from (\ref{Dpi1acting})
by inverting $a_n.$
First, we claim that
\beq \label{p1p2}
\hat D_{\pi_1} (\mathbf{a})  \mathcal{I}(\mathbf{a},\mathbf{b})  = \hat D_{\pi_2} (\mathbf{a})  \mathcal{I}(\mathbf{a},\mathbf{b}) \,.
\eeq
Second, we claim that the hypermultiplet index  $\mathcal{I}(\mathbf{a},\mathbf{b})$ intertwines the action of the $\mathrm{SO}$ and $\mathrm{USp}$
conjugated Macdonald operators,
\beq \label{intertwine}
\hat D_{\pi_1} (\mathbf{a})  \mathcal{I}(\mathbf{a},\mathbf{b})  = \hat D_{\mathrm{USp}} (\mathbf{b})  \mathcal{I}(\mathbf{a},\mathbf{b}) \,.
\eeq
While (\ref{p1p2})  and (\ref{intertwine}) are algebraic identities,
 checking them is highly tedious. For $D_2$ and $D_3,$ the different equalities were checked exactly. For higher ranks numerical evidence was obtained, up to rank $15$, by assigning random numbers smaller than one for $t$ and $q$, and random points on the unit circle for the fugacities. It would be  nice to find
an analytic proof. We also expect (but have not checked) that these identities admit the natural generalization
to generic values of the three superconformal fugacities, with Macdonald operators replaced by elliptic RS operators.

\bibliographystyle{JHEP}
\bibliography{SCIDtheories}

\end{document}